\documentclass[12pt]{article}
\usepackage[utf8]{inputenc}
\usepackage[T1]{fontenc}
\usepackage{amsmath}
\usepackage{mathtools}
\usepackage{array}
\usepackage{caption}
\usepackage{newtxtext,newtxmath}

\usepackage{graphicx}
\usepackage{longtable}
\usepackage[letterpaper,margin=1in]{geometry}

\renewenvironment{abstract}
	{\quotation}
	{\endquotation}

\date{}

\makeatletter
\renewcommand{\fnum@figure}{\textbf{Figure \thefigure}}
\renewcommand{\fnum@table}{\textbf{Table \thetable}}
\makeatother

\usepackage{scicite}

\usepackage{url}

\newcommand{\xmm}{\it XMM-Newton}

\def\scititle{
    An Intermediate-mass Black Hole Lurking in A Galactic Halo Caught Alive during Outburst
}
\title{\bfseries \boldmath \scititle}

\author{
C.-C. Jin$^{1,2,3\ast\dagger}$,
D.-Y. Li$^{1\dagger}$,
N. Jiang$^{4,5\dagger}$,
L.-X. Dai$^{6\ast}$,
H.-Q. Cheng$^{1}$,
J.-Z. Zhu$^{4,5}$,\and
C.-W. Yang$^{7,8}$,
A. Rau$^{9}$,
P. Baldini$^{9}$, 
T.-G. Wang$^{4,5,10}$
% ...
\and
\small$^\ast$Corresponding author. Email: ccjin@nao.cas.cn, lixindai@hku.hk, wmy@nao.cas.cn, chzhang@bao.ac.cn\and
\small$^\dagger$These authors contributed equally to this work.\and
\small$^{1}$National Astronomical Observatories, Chinese Academy of Sciences, Beijing 100101, China.\and
\small$^{2}$School of Astronomy and Space Science, University of Chinese Academy of Sciences, Beijing 100049, China.\and
\small$^{3}$Institute for Frontier in Astronomy and Astrophysics, Beijing Normal University, Beijing 102206, China.\and
\small$^{4}$Key Laboratory for Research in Galaxies and Cosmology of Chinese Academy of Sciences, \and \small Department of Astronomy, University of Science and Technology, Hefei, 230026, China.\and
\small$^{5}$School of Astronomy and Space Sciences, University of Science and Technology of China, Hefei, 230026, China.\and
\small$^{6}$Department of Physics, The University of Hong Kong, Pokfulam Road, Hong Kong, China.\and
\small$^{7}$Polar Research Institute of China, 451 Jinqiao Road, Pudong, Shanghai 200136, China.\and
\small$^{8}$Key Laboratory for Polar Science, MNR, Polar Research Institute of China, Shanghai, 200136, China.\and
\small$^{9}$Max-Planck-Institut für extraterrestrische Physik, Giessenbachstrasse 1, 85748 Garching, Germany.\and
\small$^{10}$Institute of Deep Space Sciences, Deep Space Exploration Laboratory, Hefei 230026, China.\and
\small$^{11}$Department of Physics, Anhui Normal University, Wuhu, Anhui, 241002, China.\and
\small$^{12}$School of Physics and Astronomy, Sun Yat-Sen University, Zhuhai, 519082, China.\and
\small$^{13}$CSST Science Center for the Guangdong-Hongkong-Macau Greater Bay Area, \and \small Sun Yat-Sen University, Zhuhai, 519082, China.\and
\small$^{14}$Universit\'e de Paris, CNRS, Astroparticule et Cosmologie, F-75013 Paris, France.\and
\small$^{15}$School of Physics and Astronomy, University of Leicester, University Road, Leicester, LE1 7RH, UK.\and
\small$^{16}$Xinjiang Astronomical Observatory, Chinese Academy of Sciences, 150 Science 1-Street, \and \small Urumqi, Xinjiang, 830011, China.\and
\small$^{17}$Laboratory for Space Research, Faculty of Science, The University of Hong Kong, \and \small Pokfulam Road, Hong Kong, China.\and
\small$^{18}$Kavli Institute for Astronomy and Astrophysics, Peking University, Beijing 100871, China.\and
\small$^{19}$Department of Astronomy, School of Physics, Peking University, Beijing 100871, China.\and
\small$^{20}$Key Laboratory of Optical Astronomy, National Astronomical Observatories, \and \small  Chinese Academy of Sciences, Beijing 100101, China.\and
\small$^{21}$Telespazio U.K. Ltd. for the European Space Agency (ESA), European Space Astronomy Centre (ESAC), \and \small Apartado 78, 28691 Villanueva de la Ca\~{n}ada, Madrid, Spain.\and
\small$^{22}$INAF-Osservatorio Astrofisico di Torino, Strada Osservatorio 20, I-10025 Pino Torinese, Italy.\and
\small$^{23}$Sydney Institute for Astronomy, School of Physics A28, The University of Sydney, Sydney, NSW 2006, Australia.\and
\small$^{24}$Nevada Center for Astrophysics, University of Nevada Las Vegas, NV 89154, USA.\and
\small$^{25}$Department of Physics and Astronomy, University of Nevada Las Vegas, NV 89154, USA.\and
\small$^{26}$Innovation Academy for Microsatellites, Chinese Academy of Sciences, Shanghai 201210, China.\and
\small$^{27}$Shanghai Institute of Technical Physics, Chinese Academy of Sciences, Shanghai 200083, China.\and
\small$^{28}$Key Laboratory of Technology on Space Energy Conversion,Technical Institute of Physics and Chemistry, \and \small CAS, Beijing 100190, China.\and
\small$^{29}$Key Laboratory of Particle Astrophysics, Institute of High Energy Physics, Chinese Academy of Sciences, \and \small Beijing 100049, China.\and
\small$^{30}$CEA Paris-Saclay, IRFU/D´epartement d’Astrophysique-AIM, 91191 Gif-sur-Yvette, France.\and
\small$^{31}$X-ray Astrophysics Laboratory, NASA Goddard Space Flight Center, Greenbelt, MD 20771, USA.\and
\small$^{32}$Cahill Center for Astronomy and Astrophysics, California Institute of Technology, Pasadena, CA 91125, USA.\and
\small$^{33}$North Night Vision Technology Co., LTD, Nanjing, China.\and
\small$^{34}$European Space Agency, ESTEC, Keplerlaan 1, NL-2200 AG, Noordwijk, The Netherlands.\and
\small$^{35}$Department of Astronomy, School of Physics, Huazhong University of Science and Technology, \and \small Wuhan, Hubei 430074, China.\and
\small$^{36}$Institute of Space Sciences (ICE), Consejo Superior de Investigaciones Científicas (CSIC), Barcelona, Spain.\and
\small$^{37}$Institut d’Estudis Espacials de Catalunya (IEEC), Barcelona, Spain.\and
\small$^{38}$National Space Science Center, Chinese Academy of Sciences, China.\and
\small$^{39}$Department of Astronomy, Xiamen University, Xiamen, 361005, China.\and
\small$^{40}$Institute of Electrical Engineering, Chinese Academy of Sciences, Beijing, 100190, China.\and
\small$^{41}$Department of Astronomy, School of Physics and Astronomy,Yunnan University, Kunming 650091, China.\and
\small$^{42}$Key Laboratory of Astroparticle Physics of Yunnan Province, Yunnan University, Kunming 650091, China.\and
\small$^{43}$Purple Mountain Observatory, Chinese Academy of Sciences, Nanjing 210023, China.
}

\begin{document} 

\maketitle

\begin{abstract} \bfseries \boldmath

Stellar-mass and supermassive black holes abound in the Universe, whereas intermediate-mass black holes (IMBHs) of $\sim 10^2-10^5$ solar masses in between are largely missing observationally, with few cases found only. Here we report the real-time discovery of a long-duration X-ray transient, EP240222a, accompanied by an optical flare with prominent H and He emission lines revealed by prompt follow-up observations. Its observed properties evidence an IMBH located unambiguously in the halo of a nearby galaxy and flaring by tidally disrupting a star---the only confirmed off-nucleus IMBH-tidal disruption event so far. This work demonstrates the potential of sensitive time-domain X-ray surveys, complemented by timely multi-wavelength follow-ups, in probing IMBHs, their environments, demographics, origins and connections to stellar-mass and supermassive black holes. 

\end{abstract}


Black holes are ubiquitous in the Universe and have profound impacts on the formation and evolution of celestial bodies. They are known to exist as stellar-mass black holes (STBHs; with masses below $\sim 100$ solar masses) in X-ray binaries and supermassive black holes (SMBHs; with masses above $\sim 10^5$ solar masses) at the centers of most galaxies. Observations show a mass gap as wide as three orders of magnitude between STBHs and SMBHs. Whether and how these two populations are connected is unknown. 
A key to this question lies in the elusive intermediate-mass black holes (IMBHs) with in-between masses, which are often considered as the seeds of SMBHs \cite{Volonteri2012, Greene2020}. 
However, IMBHs are extremely rare in observations [e.g. Omega Centauri; \cite{Haberle2024}]. This may be due partly to the difficulty in probing IMBHs based on stellar and gas dynamics \cite{Greene2020}, and partly to their possible gas-poor environments preventing them from radiating efficiently via accretion except a few, exemplified by ESO 243-49 HLX-1 \cite{Farrell2009, Godet2009, Webb2010, Soria2011}.  




When a star ventures close enough to a black hole, it can be tidally disrupted and partially accreted, resulting in a powerful electromagnetic flare\cite{Hills1975, Rees1988, komossa1999a, komossa1999b, komossa2015}. 
Such tidal disruption events (TDEs) provide an effective way to find IMBHs in dense stellar regions, such as the centers of dwarf galaxies or globular clusters, as predicted by models of IMBH formation. 
%
A progress along this line heretofore was made by the serendipitous finding in 2018 of an X-ray TDE from archival data taken in 2006–2009, 3XMM J215022.4-055108 (hereafter XMMJ2150), possibly associated with a massive star cluster at the outer skirts of a large galaxy \cite{Lin2018}. Such a physical association is hard to confirm, however, due to the lack of spectroscopic redshift measurement of that TDE since the event was only uncovered from the archival data long after the outburst. 

Clearly, timely discovery of IMBH-TDEs is essential, which can only be achieved by sensitive, wide-field surveys at high cadences from optical through ultraviolet to X-ray. The newly launched Einstein Probe [EP; \cite{Yuan2022, Yuan2025}] is a space observatory of just this kind, endowed with an unprecedented combination of large field-of-view (3850 deg$^2$) and high sensitivity in the soft X-ray band, enabled by the novel lobster-eye focusing imaging technique.

Here we report the discovery of an unrivaled case of an IMBH-TDE being positioned spectroscopically in the halo of a nearby galaxy. Its long-duration X-ray transient was initially caught alive by EP and promptly followed up with multi-wavelength observations, covering the complete cycle of the rise, peak, plateau and decay phases of the flare. These extensive temporal, spatial and spectral observations enable the establishment of the off-nucleus galactic environment for an IMBH-TDE, improving upon the previous works.

\subsection*{Discovery of a long-duration outburst in the halo of a galaxy at 137\,Mpc}
Soon after its launch, EP discovered a peculiar transient, namely EP240222a. It was first discovered on 11 March 2024 by EP as a new X-ray transient. Then by searching through archival data, earlier detections were found on 1 February 2024 by the EP pathfinder Lobster Eye Imager for Astronomy [LEIA; \cite{Zhang2022, Ling2023}], and 22 February 2024 by EP-WXT, respectively, with similar fluxes. Then an X-ray counterpart was identified in the eROSITA all-sky survey [eRASS1; \cite{Merloni2024}] on 26 May 2020, when the source flux was two orders of magnitude lower. Therefore, EP240222a is confirmed to be a peculiar long-duration X-ray transient, for which multi-wavelength follow-up observations were triggered immediately to investigate its nature.

\begin{figure}[h]%
\centering
\includegraphics[trim=0.0in -0.2in 0.0in 0.0in, clip=1, scale=0.6]{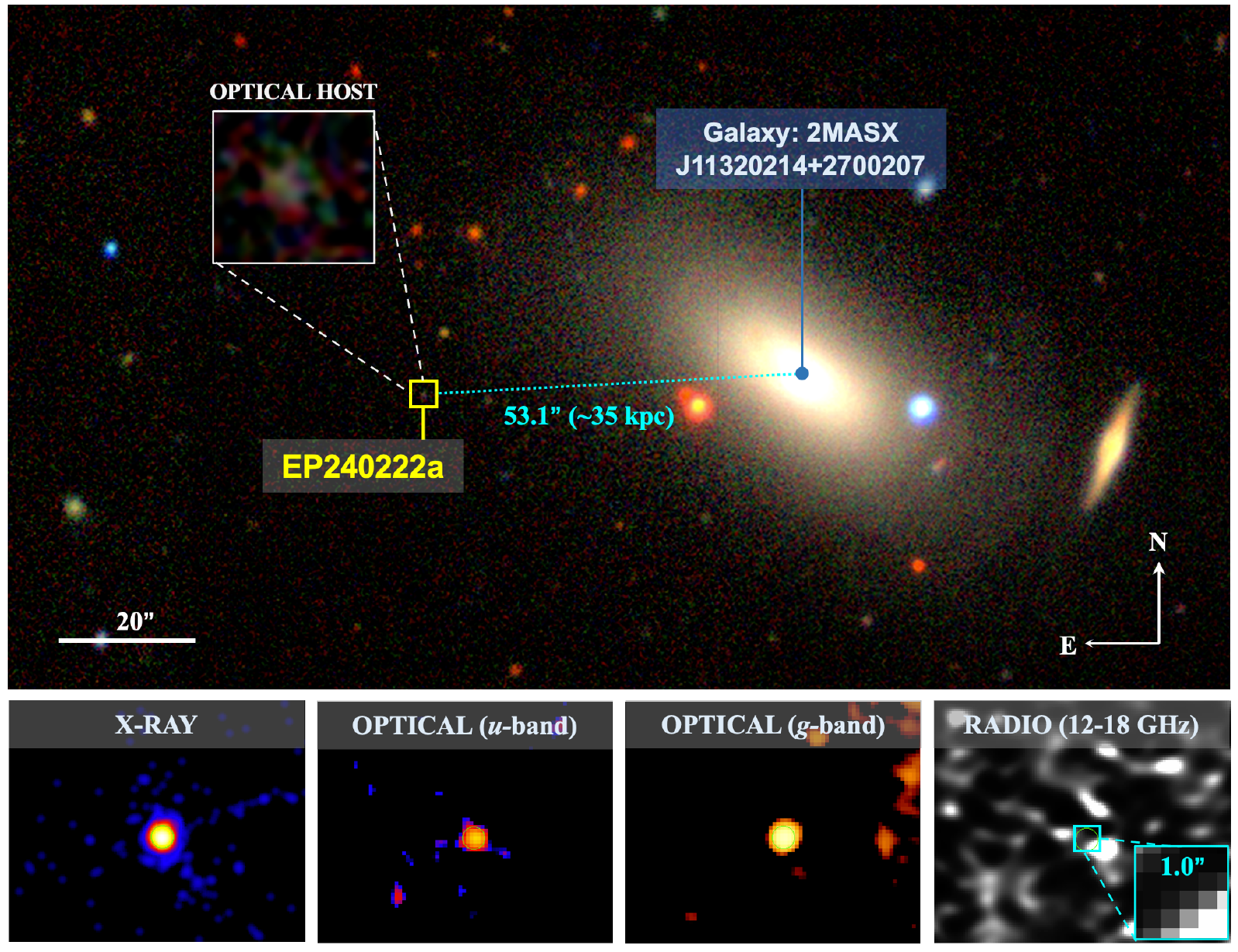}
\caption{The cosmic environment around EP240222a. The color composite image in the top panel is from the DESI Legacy Survey DR10 in $g$, $r$ and $i$ bands. The inset zoomed-in plot shows the pre-outburst optical host of EP240222a, whose magnitude is 23.98 magnitude in $g$ band. Also shown is the nearby large galaxy 2MASX J11320214$+$2700207 whose redshift is 0.03275. The angular distance between its nucleus and EP240222a is 53.1 arcsec, corresponding to a physical distance of 35 kpc at the same redshift. The set of bottom panels shows the images of EP240222a during its outburst, observed by Chandra/HRC in 0.3-10 keV (1 April 2024), WFST in the $u$ and $g$ bands (20 March 2024), and VLA in the 12-18 GHz radio band (4 April 2024), respectively. In the VLA observation, the central $1" \times 1"$ box shows the position of EP240222a, where no radio signal was detected, suggesting that the source is radio-quiet.}
\label{fig:image}
\end{figure}

Figure~\ref{fig:image} shows the cosmic environment around EP240222a and its multi-wavelength outburst images. Precise X-ray localization was provided by a 2-ks exposure with the {\it Chandra} High Resolution Camera (HRC) on 1 April 2024. The coordinates were found to be RA $=$11$^h$32$^m$06$^s$.17, Dec $= +27^\circ00^\prime17^{\prime\prime}.6$ (J2000, 1-$\sigma$ error: 0.73 arcsec). The optical counterpart was subsequently observed by a number of telescopes, including the Xinglong 2.16m Telescope, Xinglong Schmidt Telescope, the Wide Field Survey Telescope (WFST) and the Zwicky Transient Facility (ZTF), with a $g$-band brightness of $\approx$21.2 mag (see Materials and Methods). A faint pre-outburst optical host was found in the stacked image of the Dark Energy Spectroscopic Instrument (DESI) Legacy Surveys DR10, taken from March 2015 to March 2021, with a $g$-band brightness of $\approx$24.0 mag. A radio observation with the Very Large Array (VLA) on 4 April 2024 did not detect EP240222a in either C or Ku band, with a 5-$\sigma$ upper limit of 50 $\mu$Jy.

\begin{figure}[h]%
\centering
\includegraphics[trim=0.2in 0.0in 0.0in 0.0in, clip=1, scale=0.62]{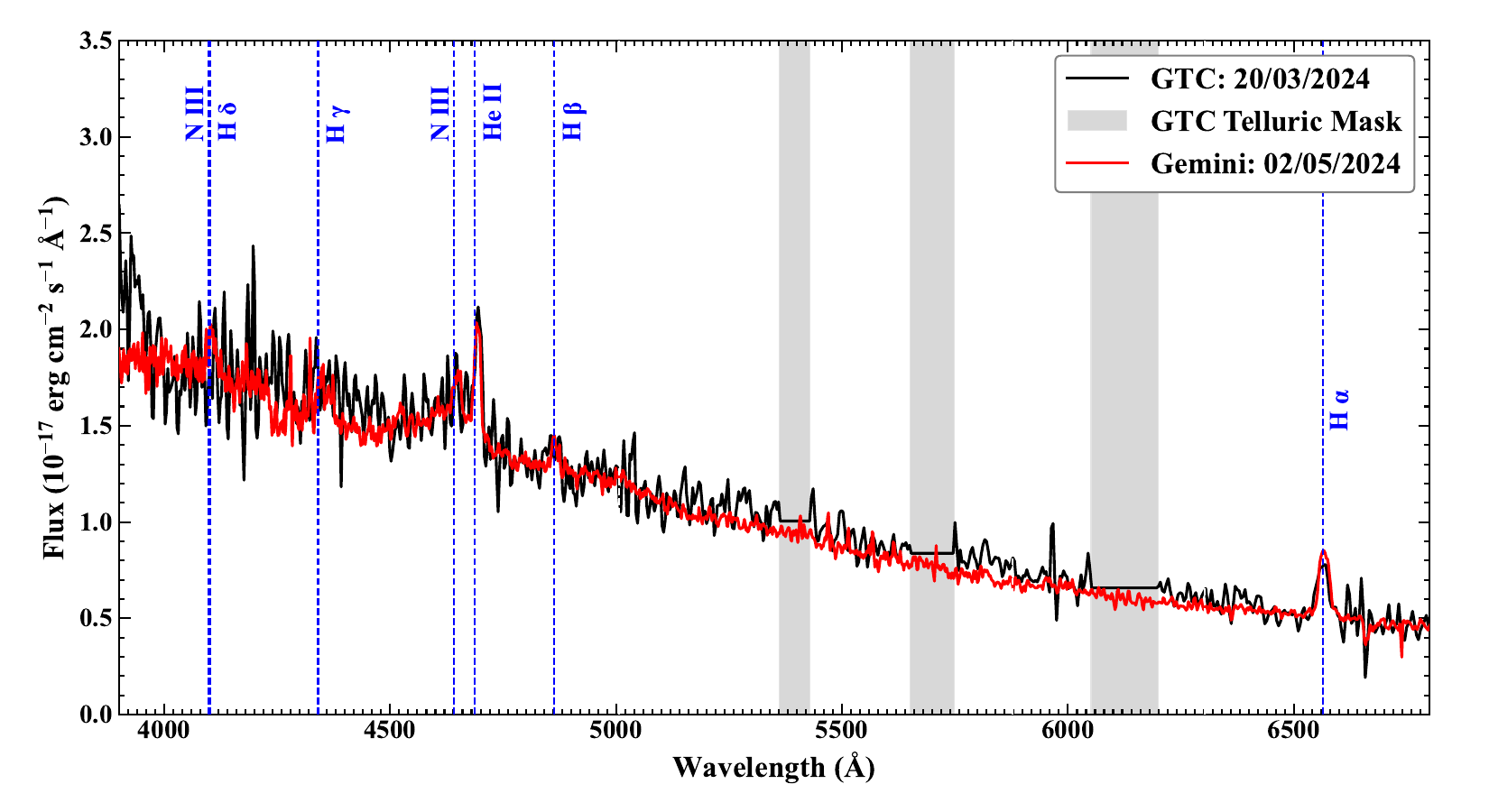} 
\caption{The optical spectra of EP240222a, taken by GTC on 20 March 2024 and Gemini on 2 May 2024. Both spectra show significant Balmer lines such as H$\alpha$ and H$\beta$, which are used to determine the redshift to be 0.032. Significant He {\sc ii} and Bowen lines like N {\sc iii} are also present, further supporting the TDE origin. The gray regions are masked out due to the telluric contamination.}
\label{fig:gtcspec1}
\end{figure}

Located 53.1 arcsec west of EP240222a is a large galaxy, namely 2MASX J11320214$+$2700207, whose redshift is $0.03275~\pm~0.00001$ \cite{Alam2015}. To ascertain if EP240222a is associated with this galaxy, we obtained its optical spectrum using the Gran Telescopio Canarias (GTC) on 20 March 2024. The GTC spectrum (Figure~\ref{fig:gtcspec1}) shows two significant emission lines, which we identify as He{\sc ii}$ \lambda$4686 and H$\alpha$, corresponding to a redshift of 0.032 $\pm$ 0.001. A later spectrum, obtained with the Gemini-North telescope on 2 May 2024, reveals additional emission lines such as H$\beta$ and the Bowen line complex N{\sc iii} $\lambda\lambda$4634, 4641, 4642, further refining the redshift to be 0.03251 $\pm$ 0.00013. The difference of this redshift from that of 2MASX J11320214 $+$2700207 is very small, corresponding to a relative velocity of only 72 $\pm$ 40 km s$^{-1}$ projected along the line-of-sight. Therefore, EP240222a is confirmed to be located in the outskirts of the galaxy 2MASX J11320214 $+$2700207, at a projected distance of $\sim$35 kpc from its nucleus.

\subsection*{A peculiar tidal disruption event}
\begin{figure*}[!h]%
\centering
\includegraphics[width=0.95\textwidth]{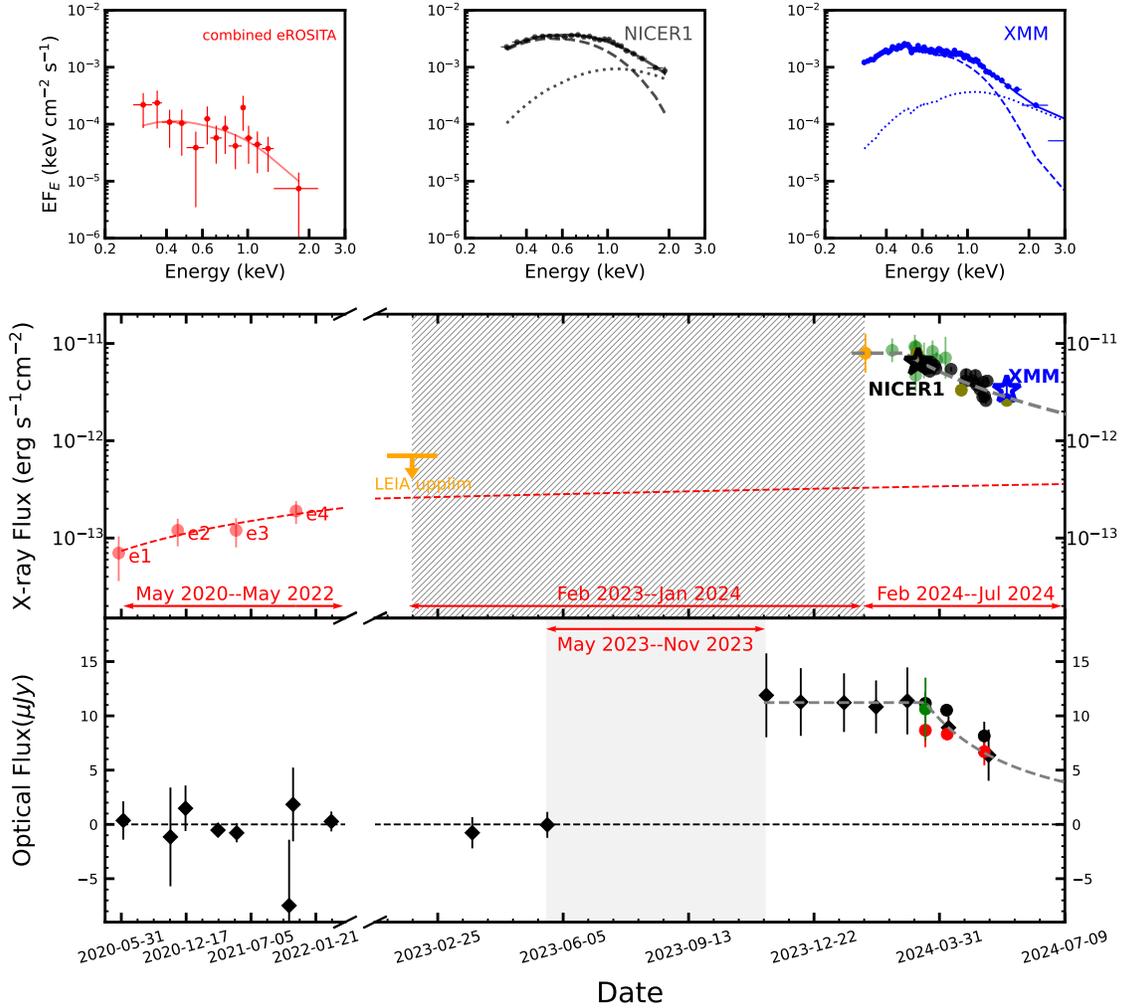}
\caption{Multi-wavelength evolution of EP240222a. {\it top panel}: X-ray spectra in different observations. The spectra in each observation can be well fitted by a standard thermal disk component (dashed lines) plus a Comptonisation component (dotted lines). {\it middle panel:} unabsorbed X-ray light curve in the 0.5--4\,keV band. The data are taken from eROSITA (red), LEIA (orange), EP-WXT (green), EP-FXT (olive), NICER (black), and {\xmm} (blue) observations. The downwards arrow shows the flux upper limit at 90\% confidence level derived from LEIA stacking data with an exposure time of 9 ks. The red dash line shows liner fit to the eROSITA detections from May 2020 to Feb. 2023, and extrapolation to June 2024. The gray dash line shows fitting to the X-ray light curve taken after Dec. 2023 with a specific function, which assumes the flux before $t_{bx}$ is constant and follows a $(t-t_{dx})^{-5/3}$ decay after $t_{bx}$.  {\it bottom panel:} optical light curves in $u$ (green), $g$ (black), $r$ (red) bands provided by ZTF (diamond) and WFST (circle). The gray dash line shows the result of fitting to the ZTF g-band light curve with a similar function to the X-ray light curve (See Materials and Methods).}
\label{fig:lc}
\end{figure*}

In order to understand the origin of EP240222a, we went through all the archival data, and kept monitoring the source with multi-wavelength facilities. Figure~\ref{fig:lc} shows the long-term X-ray light curve of EP240222a based on the observations of eROSITA, LEIA, EP, NICER and {\xmm}. After the earliest detection in the first eROSITA All-Sky Survey (eRASS1) with a 0.5--2 keV flux of $(7.4 \pm 3.4) \times 10^{-14}$\,erg\,s$^{-1}$\,cm$^{-2}$, EP240222a was subsequently detected in eRASS2, eRASS3 and eRASS4 till 24 November 2021. The X-ray flux increased gradually by a factor of three during this 1.5-year period. LEIA's coverage on February 2023 did not detect the source to an upper limit of 7.0 $\times 10^{-13}$\,erg\,s$^{-1}$\,cm$^{-2}$ in 0.5--4 keV. However, in the LEIA observation on February 2024, the X-ray flux had increased drastically to $\approx 6 \times 10^{-12}$\,erg\,s$^{-1}$\,cm$^{-2}$, and remained stable until late March 2024, then followed by a flux decay of a factor of two within the following two months. To sum up, the complete outburst of EP240222a can be described by a slow and almost linear rising phase lasting for at least two years, then a significant outburst with flux increased by more than one order of magnitude within less than one year, and then a peak plateau phase lasting for two months, and finally a slow decay phase until now. By fitting a constant plus a power law decay with an index of $-$5/3 to the post-outburst light curve, the start time of the X-ray flux decay is estimated to be around 3 March 2024 (Modified Julian Date, MJD 60372).

In all the X-ray observations, EP240222a exhibited a very soft spectrum, as shown in the top panels of Figure~\ref{fig:lc}. The spectrum can be divided into two components, below 1.5 keV it is dominated by a thermal accretion disc component with a disk temperature of $\sim$210 eV. The hard X-rays between 2--10 keV was well constrained by the deep {\xmm} observation on 24 May 2024, which showed a steep power law shape with a photon index of $3.9\pm0.2$. But this component was very weak, contributing towards only 5 percent of the total X-ray flux. This distinct X-ray spectral shape is apparently inconsistent with the typical X-ray spectra of active galactic nuclei (AGN). Also, at a distance of 136.6 Mpc, the average X-ray flux at the peak plateau phase indicates a luminosity of $1.5 \times 10^{43}$\,erg\,s$^{-1}$, effectively ruling out physical origins associated with the eruption of stellar-mass compact objects.

The bottom panel of Figure~\ref{fig:lc} shows the long-term optical light curve of EP240222a in the $g$-band, obtained with the public ZTF data and new WFST observations. Before June 2023 the source was below the detection limit of ZTF. Then in the ZTF observation on 14 November 2023, the optical brightness had increased significantly to a $g$-band magnitude of 21.28 $\pm$ 0.33 from the Legacy archival magnitude of 24, suggesting that there was a significant optical outburst within the six months between June and December 2023. Then the optical luminosity stayed in a plateau phase until late March 2024, followed by a slow decay. The luminosity in the plateau phase was $7.7 \times 10^{40}$\,erg\,s$^{-1}$, more than 2 orders of magnitude lower than that of the contemporary X-ray luminosity. Fitting the same model of the X-ray light curve to the optical, the start time of the optical decay is estimated to be around 19 March 2024 (MJD 60388).

Considering the extragalactic origin, the X-ray and optical long-term outburst, the extremely soft X-ray spectrum, the non-detection of [O {\sc iii}] $\lambda$4959/5007 (disfavoring an AGN) and the detection of strong and broad He{\sc ii}$ \lambda$4686 and Bowen N{\sc iii}, EP240222a can be classified as a TDE. Moreover, even as a TDE, EP240222a is exceptionally unique, being the first discovered in outburst with spectroscopic redshift confirmation to be occurred in the outskirts of an elliptical galaxy, strongly suggesting that it could host an IMBH. Indeed, as describe below, additional multi-wavelength follow-up observations consistently support that the black hole of EP240222a is most likely an IMBH.

\subsection*{Multi-wavelength evidence for an IMBH}

\begin{figure}[h]%
\centering
\includegraphics[width=\textwidth]{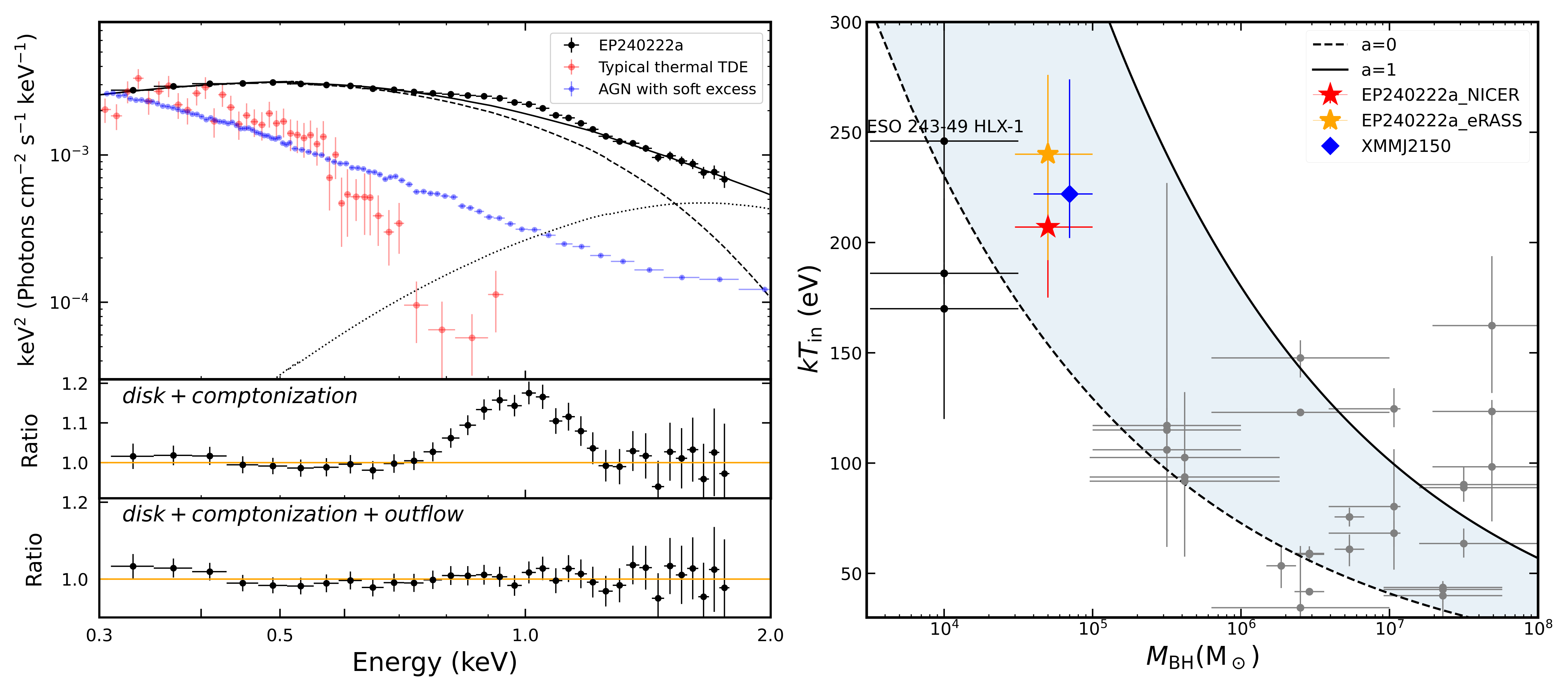}
\caption{Unique X-ray spectral properties of EP240222a. {\it left panel:} comparison between EP240222a (black circles), the thermal TDE ASASSN-14li (red circles), and the AGN RE J1034+396 (blue circles). The combined NICER spectrum of EP240222a was first fitted with a thermal disk (dash line) plus Comptonisation (dotted line) model, which leaves a significant broad excess around 1 keV, as shown by the residual in the middle panel. Then an outflow component was added, as modelled by {\tt gsmooth*xillverTDE}, which effectively removes the 1 keV excess, as shown by the residual in the bottom panel. {\it right panel:} comparison of the inner disk temperature and inferred black hole mass of EP240222a with some typical X-ray thermal TDEs in \cite{Saxton2021} (grey dots) and the Hyper-Luminous X-ray source ESO 243-49 HLX-1 (black dots). The shaded region is enclosed by the zero and maximal spin standard disk models at the Eddington mass accretion rate.}
\label{fig:xrayspec_comp}
\end{figure}

Firstly, the X-ray spectra of EP240222a measured during the peak plateau phase were modeled. Figure~\ref{fig:xrayspec_comp} (left panel) shows the combined NICER spectrum taken from 14 March to 3 May 2024, whose shape differs significantly from the soft X-ray excess of AGN \cite{Crummy2006, Jin2009, Jin2023} and the typical X-ray spectrum of thermal TDEs \cite{Kara2018, Wevers2021}. We used a model consisting of a thermal accretion disc and Comptonisation to fit the spectrum. The inner disc temperature was derived to be $T_{\rm in} \sim 210$ eV. 
%
For the standard thin disc model \cite{Shakura1973}, we have $T_{\rm in}\propto 230$ eV $(\dot{M}/\dot{M}_{\rm Edd})^{1/4} (M_{\rm BH} / 10^4 M_\odot)^{-1/4}$
for a zero-spin Schwarzschild black hole, and $T_{\rm in}\propto 570$ eV $(\dot{M}/\dot{M}_{\rm Edd})^{1/4} (M_{\rm BH} / 10^4 M_\odot)^{-1/4}$ for a maximal-spin Kerr black hole, where $\dot{M}$ is the mass accretion rate, $\dot{M}_{\rm Edd}$ the Eddington mass accretion rate and $M_{\rm BH}$ the black hole mass. Assuming EP240222a is accreting at the Eddington limit $\dot{M}=\dot{M}_{\rm Edd}$, then the derived $T_{\rm in} \sim $210 eV corresponds to a black hole mass in the range of $10^{4-5} M_{\odot}$. However, the bolometric luminosity of EP240222a during the peak plateau phase was $1.5 \times 10^{43}$ erg s$^{-1}$, which significantly exceeds the Eddington limit for $M_{\rm BH}=10^{4-5} M_{\odot}$. Considering that the radiative efficiency of super-Eddington accretion flows is lower than that of a standard disc \cite{Poutanen2007, Jiang2014, Jin2023}, the inferred black hole mass should be even smaller, so as to shift the Wein tail of the inner-disc blackbody to higher energies to match the observed X-ray spectra. 

Furthermore, a significant broad feature at $\sim$1 keV was detected during the peak plateau phase. Similar features have been seen in other thermal X-ray TDEs \cite{Kara2018, Masterson2022, Yao2024}, which can be well explained by the absorption or reflection of soft X-ray photons in winds with speeds faster than $0.1c$ (e.g., using the {\tt xillverTDE} model, see Materials and Methods). Such ultrafast, optically thick winds further support the presence of a super-Eddington accretion state \cite{Jiang2014, Sadowski2016}, and therefore a lower black hole mass. In such a super-Eddington TDE system, the X-ray emission produced from the inner disk may be reprocessed into UV and optical emission in the optically thick super-Eddington winds and disks \cite{Dai18, Thomsen22}. Indeed, our numerical simulation shows that the optical emissions of EP240222a can be produced due to reprocessing of soft X-rays as observed (see Materials and Methods). Then the small optical--to--X-ray flux ratio indicates that the disk is viewed at a small inclination from the pole \cite{Dai18}. Along this direction, there is not much reprocessing material compared to the edge-on direction, which also explains why there is no significant absorption in the X-ray spectrum.

We thus used more realistic super-Eddington accretion models for detailed modeling of the X-ray spectra of EP240222a, including {\tt slimdisk} \cite{Wen2021}, {\tt agnslim} \cite{Kubota2019} and {\tt tdediskcspec} \cite{Mummery2021, Mummery2023} in {\sc Xspec} \cite{Arnaud1996}.
These models fit the observed spectra well and all give $M_{\rm BH}$ in the range of $10^{4-5} M_{\odot}$, depending on the black hole spin (see Materials and Methods).
Specifically, a joint fit of the multi-epoch X-ray spectra with the {\tt slimdisk} model provides the best constraint on the black hole mass $M_{\rm BH}\sim  (7.7\pm4.0)\times10^4 M_\odot$ and the spin parameter $a>0.7$ at the $1\sigma$ confidence level. Fits using the {\tt agnslim} and {\tt tdediskcspec} models yield consistent $M_{\rm BH}$ values but with larger uncertainties and no constraints on the spin.
The right panel of Figure~\ref{fig:xrayspec_comp} compares the derived inner disc temperature $T_{\rm in}$ (assuming a thin disc; see above) and $M_{\rm BH}$ of EP240222a with those of some typical known thermal X-ray TDEs.
It can be seen that the $T_{\rm in}$ of EP240222a is similar to those of the other off-center TDE XMMJ2150 and the high state of ESO 243-49 HLX-1 (whose peak luminosity was one order of magnitude lower), but significantly higher than the temperatures of $\sim$ 100 eV in typical thermal X-ray TDEs \cite{komossa2015, Saxton2021}.

Secondly, the DESI Legacy Survey provides important constraints on the spectral energy distribution (SED) of the pre-outburst optical host of EP240222a in the $g$, $r$, $i$, and $z$ bands. We found a stellar mass of $\log M_*(M_\odot)= 6.45-7.26$ (90\% confidence) with a median value of 6.91 (see Materials and Methods). The blue color of $g-r \approx 0.55$ mag implies a dwarf galaxy instead of a globular cluster. The stellar mass implies a black hole mass of tens of thousands of solar masses based on the local black hole mass vs. galaxy stellar mass correlation \cite{Kormendy2013, Reines2015, Greene2020, Angus2022}. This stellar mass is also similar to that of Omega Centauri---the largest globular cluster in the Milky Way \cite{Haberle2024}---and the optical host of XMMJ2150 (assuming it has the same redshift as its nearby galaxy), both of which are suggested to contain an IMBH.

Thirdly, a closer look at the combined eROSITA spectrum of EP240222a shows that the X-ray emission during the rise phase before the plateau have retained a similar spectral shape with a temperature of 240 eV since eRASS1 in 2020 (see Materials and Methods), although the fluxes at this stage were two orders of magnitude lower than the peak. Such a thermal spectrum disfavors an AGN origin, rather suggesting that the TDE's rising phase should have started at least four years ago. Intriguingly, this rising time is the longest for any TDE known, and also much longer than the typical predicted fallback time of about 10 days for a solar-type star TDE around a $10^5 M_\odot$ MBH \cite{Evans1989, Chen2018}. 

For this exceptionally long rise timescale, one possible explanation is that it results from the accretion disk requiring an extended period to form and circularize via stream self-collisions. this extended timescale is due to the disk needing to form and circularize over a long period through stream self-collisions. Using a first-order calculation on the disk formation efficiency \cite{Dai2015, Wong2022}, we find that the disk formation time of $> 4$ years observed in EP240222a is consistent with the scenario of a solar-type star being tidally disrupted by a $< 10^5 M_\odot$ black hole on an orbit with a penetration parameter $\beta\sim1$, although some parameter degeneracy exists (see Materials and Methods). The sudden increase of the X-ray flux after 2023 indicates that the disk formation process has accelerated since then, which might be caused by the interaction between the debris stream and a small disk already formed at this stage \cite{Huang24}. Of course, given our limited knowledge of the properties of IMBH accretion systems, we cannot rule out the possibility that EP240222a might belong to another type of unique TDE system, such as an IMBH-white dwarf system, though previous theoretical studies suggest that such systems would exhibit shorter timescales \cite{MacLeod2016, Maguire2020}.

\subsection*{Prospects}

EP240222a was the first EP-TDE, detected just several weeks after the first light of EP-WXT, implying such events being not too scarce. The event rate is estimated to be $3.2 \times 10^{-7}$ gal$^{-1}$ yr$^{-1}$ for a local galaxy density of $1.4\times10^{-2}$ Mpc$^{-3}$ \cite{Donley2002} (see Materials and Methods), consistent with that estimated from XMMJ2150 \cite{Lin2018} and theoretical predictions by \cite{Chang24} as well. This discovery may signify promising prospects for detecting more IMBHs with EP and other wide-field time-domain surveys, starting to populate the desolate mass gap between STBHs and SMBHs. These IMBHs are most likely the relics of the immature seeds of SMBHs. 

At least three scenarios have been proposed for the seeds of SMBHs [e.g. \cite{Netzer2013} and references therein]: (1) $\sim 10^2 M_{\odot}$ seeds as the remnants of the first population III stars; (2) $\sim 10^3 M_{\odot}$ seeds formed via the dynamical processes in dense stellar systems; and (3) $\sim 10^{4-5} M_{\odot}$ seeds formed by the direct collapse of very massive clouds of dense and metal-poor gas postulated at the centers of proto-galaxies. These scenarios assume the different environments and predict distinctive mass functions, which could be tested by the environmental and demographic studies of IMBHs. Further studies of IMBHs, such as constraining their demographics at higher redshifts through a large sample of TDEs, will eventually reveal their physical origins and their connections with stellar-mass and supermassive black holes.

\clearpage 


\bibliography{ep240222a_main_IMBH_v2} 

\begin{thebibliography}{100}
\providecommand{\url}[1]{\texttt{#1}}
\expandafter\ifx\csname urlstyle\endcsname\relax
  \providecommand{\doi}[1]{doi:\discretionary{}{}{}#1}\else
  \providecommand{\doi}{doi:\discretionary{}{}{}\begingroup
  \urlstyle{rm}\Url}\fi

\bibitem{Volonteri2012}
M.~{Volonteri}, {The Formation and Evolution of Massive Black Holes}.
  \emph{Science} \textbf{337}~(6094), 544 (2012),
  \doi{10.1126/science.1220843}.

\bibitem{Greene2020}
J.~E. {Greene}, J.~{Strader}, L.~C. {Ho}, {Intermediate-Mass Black Holes}.
  \emph{\araa} \textbf{58}, 257--312 (2020),
  \doi{10.1146/annurev-astro-032620-021835}.

\bibitem{Haberle2024}
M.~{H{\"a}berle}, \emph{et~al.}, {Fast-moving stars around an intermediate-mass
  black hole in Omega Centauri}. \emph{arXiv e-prints} arXiv:2405.06015 (2024),
  \doi{10.48550/arXiv.2405.06015}.

\bibitem{Farrell2009}
S.~A. {Farrell}, N.~A. {Webb}, D.~{Barret}, O.~{Godet}, J.~M. {Rodrigues}, {An
  intermediate-mass black hole of over 500 solar masses in the galaxy
  ESO243-49}. \emph{\nat} \textbf{460}~(7251), 73--75 (2009),
  \doi{10.1038/nature08083}.

\bibitem{Godet2009}
O.~{Godet}, D.~{Barret}, N.~A. {Webb}, S.~A. {Farrell}, N.~{Gehrels}, {First
  Evidence for Spectral State Transitions in the ESO 243-49 Hyperluminous X-Ray
  Source HLX-1}. \emph{\apjl} \textbf{705}~(2), L109--L112 (2009),
  \doi{10.1088/0004-637X/705/2/L109}.

\bibitem{Webb2010}
N.~A. {Webb}, \emph{et~al.}, {Chandra and Swift Follow-up Observations of the
  Intermediate-mass Black Hole in ESO 243-49}. \emph{\apjl} \textbf{712}~(1),
  L107--L110 (2010), \doi{10.1088/2041-8205/712/1/L107}.

\bibitem{Soria2011}
R.~{Soria}, L.~{Zampieri}, S.~{Zane}, K.~{Wu}, {X-ray study of HLX1:
  intermediate-mass black hole or foreground neutron star?} \emph{\mnras}
  \textbf{410}~(3), 1886--1894 (2011), \doi{10.1111/j.1365-2966.2010.17572.x}.

\bibitem{Hills1975}
J.~G. {Hills}, {Possible power source of Seyfert galaxies and QSOs}.
  \emph{\nat} \textbf{254}~(5498), 295--298 (1975), \doi{10.1038/254295a0}.

\bibitem{Rees1988}
M.~J. {Rees}, {Tidal disruption of stars by black holes of 10 to the 6th-10 to
  the 8th solar masses in nearby galaxies}. \emph{\nat} \textbf{333}, 523--528
  (1988), \doi{10.1038/333523a0}.

\bibitem{komossa1999a}
S.~{Komossa}, N.~{Bade}, {The giant X-ray outbursts in NGC 5905 and IC 3599:()
  hfill Follow-up observations and outburst scenarios}. \emph{\aap}
  \textbf{343}, 775--787 (1999), \doi{10.48550/arXiv.astro-ph/9901141}.

\bibitem{komossa1999b}
S.~{Komossa}, J.~{Greiner}, {Discovery of a giant and luminous X-ray outburst
  from the optically inactive galaxy pair RX J1242.6-1119}. \emph{\aap}
  \textbf{349}, L45--L48 (1999), \doi{10.48550/arXiv.astro-ph/9908216}.

\bibitem{komossa2015}
S.~{Komossa}, {Tidal disruption of stars by supermassive black holes: Status of
  observations}. \emph{Journal of High Energy Astrophysics} \textbf{7},
  148--157 (2015), \doi{10.1016/j.jheap.2015.04.006}.

\bibitem{Lin2018}
D.~{Lin}, \emph{et~al.}, {A luminous X-ray outburst from an intermediate-mass
  black hole in an off-centre star cluster}. \emph{Nature Astronomy}
  \textbf{2}, 656--661 (2018), \doi{10.1038/s41550-018-0493-1}.

\bibitem{Yuan2022}
W.~{Yuan}, C.~{Zhang}, Y.~{Chen}, Z.~{Ling}, {The Einstein Probe Mission}, in
  \emph{Handbook of X-ray and Gamma-ray Astrophysics} (Springer), p.~86 (2022),
  \doi{10.1007/978-981-16-4544-0_151-1}.

\bibitem{Yuan2025}
W.~{Yuan}, \emph{et~al.}, {Science objectives of the Einstein Probe mission}.
  \emph{arXiv e-prints} arXiv:2501.07362 (2025),
  \doi{10.48550/arXiv.2501.07362}.

\bibitem{Zhang2022}
C.~{Zhang}, \emph{et~al.}, {First Wide Field-of-view X-Ray Observations by a
  Lobster-eye Focusing Telescope in Orbit}. \emph{\apjl} \textbf{941}~(1), L2
  (2022), \doi{10.3847/2041-8213/aca32f}.

\bibitem{Ling2023}
Z.~X. {Ling}, \emph{et~al.}, {The Lobster Eye Imager for Astronomy Onboard the
  SATech-01 Satellite}. \emph{Research in Astronomy and Astrophysics}
  \textbf{23}~(9), 095007 (2023), \doi{10.1088/1674-4527/acd593}.

\bibitem{Merloni2024}
A.~{Merloni}, \emph{et~al.}, {The SRG/eROSITA all-sky survey. First X-ray
  catalogues and data release of the western Galactic hemisphere}. \emph{\aap}
  \textbf{682}, A34 (2024), \doi{10.1051/0004-6361/202347165}.

\bibitem{Alam2015}
S.~{Alam}, \emph{et~al.}, {The Eleventh and Twelfth Data Releases of the Sloan
  Digital Sky Survey: Final Data from SDSS-III}. \emph{\apjs} \textbf{219}~(1),
  12 (2015), \doi{10.1088/0067-0049/219/1/12}.

\bibitem{Crummy2006}
J.~{Crummy}, A.~C. {Fabian}, L.~{Gallo}, R.~R. {Ross}, {An explanation for the
  soft X-ray excess in active galactic nuclei}. \emph{\mnras} \textbf{365}~(4),
  1067--1081 (2006), \doi{10.1111/j.1365-2966.2005.09844.x}.

\bibitem{Jin2009}
C.~{Jin}, C.~{Done}, M.~{Ward}, M.~{Gierli{\'n}ski}, J.~{Mullaney}, {The
  Seyfert AGN RX J0136.9-3510 and the spectral state of super Eddington
  accretion flows}. \emph{\mnras} \textbf{398}~(1), L16--L20 (2009),
  \doi{10.1111/j.1745-3933.2009.00697.x}.

\bibitem{Jin2023}
C.~{Jin}, \emph{et~al.}, {The extreme super-eddington NLS1 RX J0134.2-4258 -
  II. A weak-line Seyfert linking to the weak-line quasar}. \emph{\mnras}
  \textbf{518}~(4), 6065--6082 (2023), \doi{10.1093/mnras/stac3513}.

\bibitem{Kara2018}
E.~{Kara}, L.~{Dai}, C.~S. {Reynolds}, T.~{Kallman}, {Ultrafast outflow in
  tidal disruption event ASASSN-14li}. \emph{\mnras} \textbf{474}~(3),
  3593--3598 (2018), \doi{10.1093/mnras/stx3004}.

\bibitem{Wevers2021}
T.~{Wevers}, \emph{et~al.}, {Rapid Accretion State Transitions following the
  Tidal Disruption Event AT2018fyk}. \emph{\apj} \textbf{912}~(2), 151 (2021),
  \doi{10.3847/1538-4357/abf5e2}.

\bibitem{Shakura1973}
N.~I. {Shakura}, R.~A. {Sunyaev}, {Black holes in binary systems. Observational
  appearance.} \emph{\aap} \textbf{24}, 337--355 (1973).

\bibitem{Poutanen2007}
J.~{Poutanen}, G.~{Lipunova}, S.~{Fabrika}, A.~G. {Butkevich}, P.~{Abolmasov},
  {Supercritically accreting stellar mass black holes as ultraluminous X-ray
  sources}. \emph{\mnras} \textbf{377}~(3), 1187--1194 (2007),
  \doi{10.1111/j.1365-2966.2007.11668.x}.

\bibitem{Jiang2014}
Y.-F. {Jiang}, J.~M. {Stone}, S.~W. {Davis}, {A Global Three-dimensional
  Radiation Magneto-hydrodynamic Simulation of Super-Eddington Accretion
  Disks}. \emph{\apj} \textbf{796}~(2), 106 (2014),
  \doi{10.1088/0004-637X/796/2/106}.

\bibitem{Masterson2022}
M.~{Masterson}, \emph{et~al.}, {Evolution of a Relativistic Outflow and X-Ray
  Corona in the Extreme Changing-look AGN 1ES 1927+654}. \emph{\apj}
  \textbf{934}~(1), 35 (2022), \doi{10.3847/1538-4357/ac76c0}.

\bibitem{Yao2024}
Y.~{Yao}, \emph{et~al.}, {Sub-relativistic Outflow and Hours-Timescale
  Large-amplitude X-ray Dips during Super-Eddington Accretion onto a Low-mass
  Massive Black Hole in the Tidal Disruption Event AT2022lri}. \emph{arXiv
  e-prints} arXiv:2405.11343 (2024), \doi{10.48550/arXiv.2405.11343}.

\bibitem{Sadowski2016}
A.~{S{\k{a}}dowski}, R.~{Narayan}, {Three-dimensional simulations of
  supercritical black hole accretion discs - luminosities, photon trapping and
  variability}. \emph{\mnras} \textbf{456}~(4), 3929--3947 (2016),
  \doi{10.1093/mnras/stv2941}.

\bibitem{Dai18}
L.~{Dai}, J.~C. {McKinney}, N.~{Roth}, E.~{Ramirez-Ruiz}, M.~C. {Miller}, {A
  Unified Model for Tidal Disruption Events}. \emph{\apjl} \textbf{859}~(2),
  L20 (2018), \doi{10.3847/2041-8213/aab429}.

\bibitem{Thomsen22}
L.~L. {Thomsen}, \emph{et~al.}, {Dynamical Unification of Tidal Disruption
  Events}. \emph{\apjl} \textbf{937}~(2), L28 (2022),
  \doi{10.3847/2041-8213/ac911f}.

\bibitem{Wen2021}
S.~Wen, P.~G. Jonker, N.~C. Stone, A.~I. Zabludoff, Mass, Spin, and Ultralight
  Boson Constraints from the Intermediate-mass Black Hole in the Tidal
  Disruption Event 3XMM J215022.4{\textendash}055108. \emph{\apj}
  \textbf{918}~(2), 46 (2021), \doi{10.3847/1538-4357/ac00b5}.

\bibitem{Kubota2019}
A.~{Kubota}, C.~{Done}, {Modelling the spectral energy distribution of
  super-Eddington quasars}. \emph{\mnras} \textbf{489}~(1), 524--533 (2019),
  \doi{10.1093/mnras/stz2140}.

\bibitem{Mummery2021}
A.~Mummery, {Tidal disruption event discs are larger than they seem: removing
  systematic biases in TDE X-ray spectral modelling}. \emph{Monthly Notices of
  the Royal Astronomical Society: Letters} \textbf{507}~(1), L24--L28 (2021),
  \doi{10.1093/mnrasl/slab088}.

\bibitem{Mummery2023}
A.~{Mummery}, T.~{Wevers}, R.~{Saxton}, D.~{Pasham}, {From X-rays to physical
  parameters: a comprehensive analysis of thermal tidal disruption event X-ray
  spectra}. \emph{\mnras} \textbf{519}~(4), 5828--5847 (2023),
  \doi{10.1093/mnras/stac3798}.

\bibitem{Arnaud1996}
K.~A. {Arnaud}, {XSPEC: The First Ten Years}, in \emph{Astronomical Data
  Analysis Software and Systems V}, G.~H. {Jacoby}, J.~{Barnes}, Eds., vol. 101
  of \emph{Astronomical Society of the Pacific Conference Series} (1996),
  p.~17.

\bibitem{Saxton2021}
R.~{Saxton}, S.~{Komossa}, K.~{Auchettl}, P.~G. {Jonker}, {Correction to: X-Ray
  Properties of TDEs}, Space Science Reviews, Volume 217, Issue 1, article
  id.18 (2021), \doi{10.1007/s11214-020-00759-7}.

\bibitem{Kormendy2013}
J.~{Kormendy}, L.~C. {Ho}, {Coevolution (Or Not) of Supermassive Black Holes
  and Host Galaxies}. \emph{\araa} \textbf{51}~(1), 511--653 (2013),
  \doi{10.1146/annurev-astro-082708-101811}.

\bibitem{Reines2015}
A.~E. {Reines}, M.~{Volonteri}, {Relations between Central Black Hole Mass and
  Total Galaxy Stellar Mass in the Local Universe}. \emph{\apj}
  \textbf{813}~(2), 82 (2015), \doi{10.1088/0004-637X/813/2/82}.

\bibitem{Angus2022}
C.~R. {Angus}, \emph{et~al.}, {A fast-rising tidal disruption event from a
  candidate intermediate-mass black hole}. \emph{Nature Astronomy} \textbf{6},
  1452--1463 (2022), \doi{10.1038/s41550-022-01811-y}.

\bibitem{Evans1989}
C.~R. {Evans}, C.~S. {Kochanek}, {The Tidal Disruption of a Star by a Massive
  Black Hole}. \emph{\apjl} \textbf{346}, L13 (1989), \doi{10.1086/185567}.

\bibitem{Chen2018}
J.-H. {Chen}, R.-F. {Shen}, {Tidal Disruption of a Main-sequence Star by an
  Intermediate-mass Black Hole: A Bright Decade}. \emph{\apj} \textbf{867}~(1),
  20 (2018), \doi{10.3847/1538-4357/aadfda}.

\bibitem{Dai2015}
L.~{Dai}, J.~C. {McKinney}, M.~C. {Miller}, {Soft X-Ray Temperature Tidal
  Disruption Events from Stars on Deep Plunging Orbits}. \emph{\apjl}
  \textbf{812}~(2), L39 (2015), \doi{10.1088/2041-8205/812/2/L39}.

\bibitem{Wong2022}
T.~H.~T. {Wong}, H.~{Pfister}, L.~{Dai}, {Revisiting the Rates and Demographics
  of Tidal Disruption Events: Effects of the Disk Formation Efficiency}.
  \emph{\apjl} \textbf{927}~(1), L19 (2022), \doi{10.3847/2041-8213/ac5823}.

\bibitem{Huang24}
X.~{Huang}, S.~W. {Davis}, Y.-f. {Jiang}, {Pre-peak Emission in Tidal
  Disruption Events}. \emph{arXiv e-prints} arXiv:2404.18446 (2024),
  \doi{10.48550/arXiv.2404.18446}.

\bibitem{MacLeod2016}
M.~{MacLeod}, J.~{Guillochon}, E.~{Ramirez-Ruiz}, D.~{Kasen}, S.~{Rosswog},
  {Optical Thermonuclear Transients from Tidal Compression of White Dwarfs as
  Tracers of the Low End of the Massive Black Hole Mass Function}. \emph{\apj}
  \textbf{819}~(1), 3 (2016), \doi{10.3847/0004-637X/819/1/3}.

\bibitem{Maguire2020}
K.~{Maguire}, M.~{Eracleous}, P.~G. {Jonker}, M.~{MacLeod}, S.~{Rosswog},
  {Tidal Disruptions of White Dwarfs: Theoretical Models and Observational
  Prospects}. \emph{\ssr} \textbf{216}~(3), 39 (2020),
  \doi{10.1007/s11214-020-00661-2}.

\bibitem{Donley2002}
J.~L. {Donley}, W.~N. {Brandt}, M.~{Eracleous}, T.~{Boller}, {Large-Amplitude
  X-Ray Outbursts from Galactic Nuclei: A Systematic Survey using ROSAT
  Archival Data}. \emph{\aj} \textbf{124}~(3), 1308--1321 (2002),
  \doi{10.1086/342280}.

\bibitem{Chang24}
J.~N.~Y. {Chang}, L.~{Dai}, H.~{Pfister}, R.~K. {Chowdhury}, P.~{Natarajan},
  {Rates of Stellar Tidal Disruption Events Around Intermediate-Mass Black
  Holes}. \emph{arXiv e-prints} arXiv:2407.09339 (2024),
  \doi{10.48550/arXiv.2407.09339}.

\bibitem{Netzer2013}
H.~{Netzer}, \emph{{The Physics and Evolution of Active Galactic Nuclei}}
  (2013).

\bibitem{Chen2020}
Y.~{Chen}, \emph{et~al.}, {Status of the follow-up x-ray telescope onboard the
  Einstein Probe satellite}, in \emph{Space Telescopes and Instrumentation
  2020: Ultraviolet to Gamma Ray}, J.-W.~A. {den Herder}, S.~{Nikzad},
  K.~{Nakazawa}, Eds., vol. 11444 of \emph{Society of Photo-Optical
  Instrumentation Engineers (SPIE) Conference Series} (2020), p. 114445B,
  \doi{10.1117/12.2562311}.

\bibitem{Fruscione2006}
A.~{Fruscione}, \emph{et~al.}, {CIAO: Chandra's data analysis system}, in
  \emph{Observatory Operations: Strategies, Processes, and Systems}, D.~R.
  {Silva}, R.~E. {Doxsey}, Eds., vol. 6270 of \emph{Society of Photo-Optical
  Instrumentation Engineers (SPIE) Conference Series} (2006), p. 62701V,
  \doi{10.1117/12.671760}.

\bibitem{Gabriel2004}
C.~{Gabriel}, \emph{et~al.}, {The XMM-Newton SAS - Distributed Development and
  Maintenance of a Large Science Analysis System: A Critical Analysis}, in
  \emph{Astronomical Data Analysis Software and Systems (ADASS) XIII},
  F.~{Ochsenbein}, M.~G. {Allen}, D.~{Egret}, Eds., vol. 314 of
  \emph{Astronomical Society of the Pacific Conference Series} (2004), p. 759.

\bibitem{Predehl21}
P.~{Predehl}, \emph{et~al.}, {The eROSITA X-ray telescope on SRG}. \emph{\aap}
  \textbf{647}, A1 (2021), \doi{10.1051/0004-6361/202039313}.

\bibitem{Sunyaev21}
R.~{Sunyaev}, \emph{et~al.}, {SRG X-ray orbital observatory. Its telescopes and
  first scientific results}. \emph{\aap} \textbf{656}, A132 (2021),
  \doi{10.1051/0004-6361/202141179}.

\bibitem{Brunner22}
H.~{Brunner}, \emph{et~al.}, {The eROSITA Final Equatorial Depth Survey
  (eFEDS). X-ray catalogue}. \emph{\aap} \textbf{661}, A1 (2022),
  \doi{10.1051/0004-6361/202141266}.

\bibitem{Liu22}
T.~{Liu}, \emph{et~al.}, {The eROSITA Final Equatorial-Depth Survey (eFEDS).
  The AGN catalog and its X-ray spectral properties}. \emph{\aap} \textbf{661},
  A5 (2022), \doi{10.1051/0004-6361/202141643}.

\bibitem{Buchner14}
J.~{Buchner}, \emph{et~al.}, {X-ray spectral modelling of the AGN obscuring
  region in the CDFS: Bayesian model selection and catalogue}. \emph{\aap}
  \textbf{564}, A125 (2014), \doi{10.1051/0004-6361/201322971}.

\bibitem{Buchner21}
J.~{Buchner}, {UltraNest - a robust, general purpose Bayesian inference
  engine}. \emph{The Journal of Open Source Software} \textbf{6}~(60), 3001
  (2021), \doi{10.21105/joss.03001}.

\bibitem{Fruscione06}
A.~{Fruscione}, \emph{et~al.}, {CIAO: Chandra's data analysis system}, in
  \emph{Observatory Operations: Strategies, Processes, and Systems}, D.~R.
  {Silva}, R.~E. {Doxsey}, Eds., vol. 6270 of \emph{Society of Photo-Optical
  Instrumentation Engineers (SPIE) Conference Series} (2006), p. 62701V,
  \doi{10.1117/12.671760}.

\bibitem{Simmonds18}
C.~{Simmonds}, J.~{Buchner}, M.~{Salvato}, L.~T. {Hsu}, F.~E. {Bauer}, {XZ:
  Deriving redshifts from X-ray spectra of obscured AGN}. \emph{\aap}
  \textbf{618}, A66 (2018), \doi{10.1051/0004-6361/201833412}.

\bibitem{Wilms2000}
J.~{Wilms}, A.~{Allen}, R.~{McCray}, {On the Absorption of X-Rays in the
  Interstellar Medium}. \emph{\apj} \textbf{542}~(2), 914--924 (2000),
  \doi{10.1086/317016}.

\bibitem{willingale2013}
R.~{Willingale}, R.~L.~C. {Starling}, A.~P. {Beardmore}, N.~R. {Tanvir}, P.~T.
  {O'Brien}, {Calibration of X-ray absorption in our Galaxy}. \emph{\mnras}
  \textbf{431}~(1), 394--404 (2013), \doi{10.1093/mnras/stt175}.

\bibitem{Pounds1995}
K.~A. {Pounds}, C.~{Done}, J.~P. {Osborne}, {RE 1034+39: a high-state Seyfert
  galaxy?} \emph{\mnras} \textbf{277}~(1), L5--L10 (1995),
  \doi{10.1093/mnras/277.1.L5}.

\bibitem{Titarchuk1994}
L.~{Titarchuk}, {Generalized Comptonization Models and Application to the
  Recent High-Energy Observations}. \emph{\apj} \textbf{434}, 570 (1994),
  \doi{10.1086/174760}.

\bibitem{Garcia10}
J.~{Garc{\'\i}a}, T.~R. {Kallman}, {X-ray Reflected Spectra from Accretion Disk
  Models. I. Constant Density Atmospheres}. \emph{\apj} \textbf{718}~(2),
  695--706 (2010), \doi{10.1088/0004-637X/718/2/695}.

\bibitem{Thomsen19}
L.~L. {Thomsen}, J.~{Lixin Dai}, E.~{Ramirez-Ruiz}, E.~{Kara}, C.~{Reynolds},
  {X-Ray Fluorescence from Super-Eddington Accreting Black Holes}. \emph{\apjl}
  \textbf{884}~(1), L21 (2019), \doi{10.3847/2041-8213/ab4518}.

\bibitem{Zhang24}
Z.~{Zhang}, \emph{et~al.}, {Modeling X-Ray Multi-Reflection in Super-Eddington
  Winds}. \emph{arXiv e-prints} arXiv:2407.08596 (2024),
  \doi{10.48550/arXiv.2407.08596}.

\bibitem{Wen2022}
S.~{Wen}, P.~G. {Jonker}, N.~C. {Stone}, S.~{Van Velzen}, A.~I. {Zabludoff},
  {Optical/UV emission in the Tidal Disruption Event ASASSN-14li: implications
  of disc modelling}. \emph{\mnras} \textbf{522}~(1), 1155--1168 (2023),
  \doi{10.1093/mnras/stad991}.

\bibitem{Cao2023}
Z.~{Cao}, P.~G. {Jonker}, S.~{Wen}, N.~C. {Stone}, A.~I. {Zabludoff}, {The
  rapidly spinning intermediate-mass black hole 3XMM J150052.0+015452}.
  \emph{\mnras} \textbf{519}~(2), 2375--2390 (2023),
  \doi{10.1093/mnras/stac3539}.

\bibitem{Wen2020}
S.~{Wen}, P.~G. {Jonker}, N.~C. {Stone}, A.~I. {Zabludoff}, D.~{Psaltis},
  {Continuum-fitting the X-Ray Spectra of Tidal Disruption Events}. \emph{\apj}
  \textbf{897}~(1), 80 (2020), \doi{10.3847/1538-4357/ab9817}.

\bibitem{Zdziarski_2020}
A.~A. {Zdziarski}, M.~{Szanecki}, J.~{Poutanen}, M.~{Gierli{\'n}ski},
  P.~{Biernacki}, {Spectral and temporal properties of Compton scattering by
  mildly relativistic thermal electrons}. \emph{\mnras} \textbf{492}~(4),
  5234--5246 (2020), \doi{10.1093/mnras/staa159}.

\bibitem{Bertin2010}
E.~{Bertin}, {SWarp: Resampling and Co-adding FITS Images Together},
  Astrophysics Source Code Library, record ascl:1010.068 (2010).

\bibitem{AUTOPHOT}
S.~J. {Brennan}, M.~{ Fraser}, {The Automated Photometry of Transients pipeline
  (AUTOPHOT)}. \emph{\aap} \textbf{667}, A62 (2022),
  \doi{10.1051/0004-6361/202243067}.

\bibitem{Flewelling2020}
H.~A. {Flewelling}, \emph{et~al.}, {The Pan-STARRS1 Database and Data
  Products}. \emph{\apjs} \textbf{251}~(1), 7 (2020),
  \doi{10.3847/1538-4365/abb82d}.

\bibitem{Fan2016}
Z.~{Fan}, \emph{et~al.}, {The Xinglong 2.16-m Telescope: Current Instruments
  and Scientific Projects}. \emph{\pasp} \textbf{128}~(969), 115005 (2016),
  \doi{10.1088/1538-3873/128/969/115005}.

\bibitem{Kron1980}
R.~G. {Kron}, {Photometry of a complete sample of faint galaxies.} \emph{\apjs}
  \textbf{43}, 305--325 (1980), \doi{10.1086/190669}.

\bibitem{WangT2023}
T.~{Wang}, \emph{et~al.}, {Science with the 2.5-meter Wide Field Survey
  Telescope (WFST)}. \emph{Science China Physics, Mechanics, and Astronomy}
  \textbf{66}~(10), 109512 (2023), \doi{10.1007/s11433-023-2197-5}.

\bibitem{SF2011}
E.~F. {Schlafly}, D.~P. {Finkbeiner}, {Measuring Reddening with Sloan Digital
  Sky Survey Stellar Spectra and Recalibrating SFD}. \emph{\apj}
  \textbf{737}~(2), 103 (2011), \doi{10.1088/0004-637X/737/2/103}.

\bibitem{Greiner2008}
J.~{Greiner}, \emph{et~al.}, {GROND{\textemdash}a 7-Channel Imager}.
  \emph{\pasp} \textbf{120}~(866), 405 (2008), \doi{10.1086/587032}.

\bibitem{kruehler08}
T.~{Kr{\"u}hler}, \emph{et~al.}, {The 2175 {\r{A}} Dust Feature in a Gamma-Ray
  Burst Afterglow at Redshift 2.45}. \emph{\apj} \textbf{685}~(1), 376--383
  (2008), \doi{10.1086/590240}.

\bibitem{skrutskie06}
M.~F. {Skrutskie}, \emph{et~al.}, {The Two Micron All Sky Survey (2MASS)}.
  \emph{\aj} \textbf{131}~(2), 1163--1183 (2006), \doi{10.1086/498708}.

\bibitem{Masci2023}
F.~J. {Masci}, \emph{et~al.}, {A New Forced Photometry Service for the Zwicky
  Transient Facility}. \emph{arXiv e-prints} arXiv:2305.16279 (2023),
  \doi{10.48550/arXiv.2305.16279}.

\bibitem{vV2021}
S.~{van Velzen}, \emph{et~al.}, {Seventeen Tidal Disruption Events from the
  First Half of ZTF Survey Observations: Entering a New Era of Population
  Studies}. \emph{\apj} \textbf{908}~(1), 4 (2021),
  \doi{10.3847/1538-4357/abc258}.

\bibitem{Cepa2003}
J.~{Cepa}, \emph{et~al.}, {OSIRIS tunable imager and spectrograph for the GTC.
  Instrument status}, in \emph{Instrument Design and Performance for
  Optical/Infrared Ground-based Telescopes}, M.~{Iye}, A.~F.~M. {Moorwood},
  Eds., vol. 4841 of \emph{Society of Photo-Optical Instrumentation Engineers
  (SPIE) Conference Series} (2003), pp. 1739--1749, \doi{10.1117/12.460913}.

\bibitem{Hook2004}
I.~M. {Hook}, \emph{et~al.}, {The Gemini-North Multi-Object Spectrograph:
  Performance in Imaging, Long-Slit, and Multi-Object Spectroscopic Modes}.
  \emph{\pasp} \textbf{116}~(819), 425--440 (2004), \doi{10.1086/383624}.

\bibitem{Pypeit1}
J.~X. {Prochaska}, \emph{et~al.}, {pypeit/PypeIt: Release 1.0.0} (2020),
  \doi{10.5281/zenodo.3743493}.

\bibitem{Pypeit2}
J.~{Prochaska}, \emph{et~al.}, {PypeIt: The Python Spectroscopic Data Reduction
  Pipeline}. \emph{The Journal of Open Source Software} \textbf{5}~(56), 2308
  (2020), \doi{10.21105/joss.02308}.

\bibitem{CASA2022}
{CASA Team}, \emph{et~al.}, {CASA, the Common Astronomy Software Applications
  for Radio Astronomy}. \emph{\pasp} \textbf{134}~(1041), 114501 (2022),
  \doi{10.1088/1538-3873/ac9642}.

\bibitem{Dey2019}
A.~{Dey}, \emph{et~al.}, {Overview of the DESI Legacy Imaging Surveys}.
  \emph{\aj} \textbf{157}~(5), 168 (2019), \doi{10.3847/1538-3881/ab089d}.

\bibitem{Moustakas2013}
J.~{Moustakas}, \emph{et~al.}, {PRIMUS: Constraints on Star Formation Quenching
  and Galaxy Merging, and the Evolution of the Stellar Mass Function from z =
  0-1}. \emph{\apj} \textbf{767}~(1), 50 (2013),
  \doi{10.1088/0004-637X/767/1/50}.

\bibitem{Moustakas2017}
J.~{Moustakas}, {iSEDfit: Bayesian spectral energy distribution modeling of
  galaxies}, Astrophysics Source Code Library, record ascl:1708.029 (2017).

\bibitem{CIGALE}
M.~{Boquien}, \emph{et~al.}, {CIGALE: a python Code Investigating GALaxy
  Emission}. \emph{\aap} \textbf{622}, A103 (2019),
  \doi{10.1051/0004-6361/201834156}.

\bibitem{Schutte2019}
Z.~{Schutte}, A.~E. {Reines}, J.~E. {Greene}, {The Black Hole-Bulge Mass
  Relation Including Dwarf Galaxies Hosting Active Galactic Nuclei}.
  \emph{\apj} \textbf{887}~(2), 245 (2019), \doi{10.3847/1538-4357/ab35dd}.

\bibitem{Chang2015}
Y.-Y. {Chang}, A.~{van der Wel}, E.~{da Cunha}, H.-W. {Rix}, {Stellar Masses
  and Star Formation Rates for 1M Galaxies from SDSS+WISE}. \emph{\apjs}
  \textbf{219}~(1), 8 (2015), \doi{10.1088/0067-0049/219/1/8}.

\bibitem{KH2013}
J.~{Kormendy}, L.~C. {Ho}, {Coevolution (Or Not) of Supermassive Black Holes
  and Host Galaxies}. \emph{\araa} \textbf{51}~(1), 511--653 (2013),
  \doi{10.1146/annurev-astro-082708-101811}.

\bibitem{Yang2007}
X.~{Yang}, \emph{et~al.}, {Galaxy Groups in the SDSS DR4. I. The Catalog and
  Basic Properties}. \emph{\apj} \textbf{671}~(1), 153--170 (2007),
  \doi{10.1086/522027}.

\bibitem{Cooper2011}
A.~P. {Cooper}, \emph{et~al.}, {The Formation of Shell Galaxies Similar to NGC
  7600 in the Cold Dark Matter Cosmogony}. \emph{\apjl} \textbf{743}~(1), L21
  (2011), \doi{10.1088/2041-8205/743/1/L21}.

\bibitem{Shiokawa15}
H.~{Shiokawa}, J.~H. {Krolik}, R.~M. {Cheng}, T.~{Piran}, S.~C. {Noble},
  {General Relativistic Hydrodynamic Simulation of Accretion Flow from a
  Stellar Tidal Disruption}. \emph{\apj} \textbf{804}~(2), 85 (2015),
  \doi{10.1088/0004-637X/804/2/85}.

\bibitem{Bonnerot2016}
C.~{Bonnerot}, E.~M. {Rossi}, G.~{Lodato}, D.~J. {Price}, {Disc formation from
  tidal disruptions of stars on eccentric orbits by Schwarzschild black holes}.
  \emph{\mnras} \textbf{455}~(2), 2253--2266 (2016),
  \doi{10.1093/mnras/stv2411}.

\bibitem{Dai15}
L.~{Dai}, J.~C. {McKinney}, M.~C. {Miller}, {Soft X-Ray Temperature Tidal
  Disruption Events from Stars on Deep Plunging Orbits}. \emph{\apjl}
  \textbf{812}~(2), L39 (2015), \doi{10.1088/2041-8205/812/2/L39}.

\bibitem{Wong22}
T.~H.~T. {Wong}, H.~{Pfister}, L.~{Dai}, {Revisiting the Rates and Demographics
  of Tidal Disruption Events: Effects of the Disk Formation Efficiency}.
  \emph{\apjl} \textbf{927}~(1), L19 (2022), \doi{10.3847/2041-8213/ac5823}.

\bibitem{Kasen06}
D.~{Kasen}, R.~C. {Thomas}, P.~{Nugent}, {Time-dependent Monte Carlo Radiative
  Transfer Calculations for Three-dimensional Supernova Spectra, Light Curves,
  and Polarization}. \emph{\apj} \textbf{651}, 366--380 (2006),
  \doi{10.1086/506190}.

\bibitem{Roth16}
N.~{Roth}, D.~{Kasen}, J.~{Guillochon}, E.~{Ramirez-Ruiz}, {The X-Ray through
  Optical Fluxes and Line Strengths of Tidal Disruption Events}. \emph{\apj}
  \textbf{827}, 3 (2016), \doi{10.3847/0004-637X/827/1/3}.

\end{thebibliography}
\bibliographystyle{sciencemag}

\clearpage 



\section*{Acknowledgments}
Here you can thank helpful colleagues who did not meet the journal's authorship criteria, or
provide other acknowledgements that don't fit the (compulsory) subheadings below.
Formatting requirements for each of these sections differ between the \textit{Science}-family
journals; consult the instructions to authors on the journal website for full details.
\paragraph*{Funding:}
This work is based on the data obtained with Einstein Probe, a space mission supported by Strategic Priority Program on Space Science of Chinese Academy of Sciences, in collaboration with ESA, MPE and CNES (Grant No. XDA15310000), the Strategic Priority Research Program of the Chinese Academy of Sciences (Grant No. XDB0550200), and the National Key R\&D Program of China (2022YFF0711500). CJ was supported by the National Natural Science Foundation of China (12473016). NJ and TGW were supported by the National Natural Science Foundation of China (12192221,12393814). LXD was supported by the National Science Foundation of China and the Hong Kong Research Grants Council (12122309, N\_HKU782/23, 17314822, 17304821). HQC was supported by National Natural Science Foundation of China 12203071. YQX was supported by National Key R\&D Program of China 2023YFA1608100, National Natural Science Foundation of China 12025303 and 12393814. LCH was supported by the National Science Foundation of China (11991052, 12233001), the National Key R\&D Program of China (2022YFF0503401), and the China Manned Space Project (CMS-CSST-2021-A04, CMS-CSST-2021-A06). We acknowledge the data resources and technical support provided by the China National Astronomical Data Center, the Astronomical Science Data Center of the Chinese Academy of Sciences, and the Chinese Virtual Observatory. We acknowledge the observational data taken at Chandra (PI: C.-C. Jin), {\xmm} (PI: H.-N. Yang), NICER (PI: H.-Q. Cheng), GTC (PI: C.-C. Jin), GEMINI (PI: N. Jiang), VLA (PI: X.-W. Shu). We acknowledge the support of the staff of the Xinglong 2.16m telescope. This work was partially supported by the Open Project Program of the CAS Key Laboratory of Optical Astronomy, National Astronomical Observatories, Chinese Academy of Sciences.

\paragraph*{Author contributions:}

W.Y. has been leading the Einstein Probe project as Principal Investigator since
the mission proposal phase.
C.-C.J., D.-Y.L., N.J., L.-X.D., W.Y. initiated the study. C.-C.J., L.-X.D.,D.-Y.L., N.J., H.-Q.C., T.-G.W., H.-Y.Z. coordinated the scientific investigations of the event. 
D.-Y.L., C.-C.J. processed and analysed the LEIA, EP-WXT, EP-FXT, {\xmm}, and Chandra data. H.-Q.C. processed and analysed the NICER data. P.B and A.R. analysed the eROSITA data. H.-N.Y. contributed to the {\xmm} data taking. J.-Z.Z., N.J. and T.-G.W. searched the ZTF data and contributed to the WFST data taking and analysis. C.-C.J., F.X. and J.F.L. contributed to the GTC data taking. C.-W.Y. and H.-Y.Z. analysed the GTC data. Y.-B.W. and N.J. contributed to the Gemini data acquisition and analysis. X.-W.S contributed to the radio data acquisition and analysis. T.-G.W. and J.-Z.Z. performed SED analysis of the TDE host. W.-J.Z. carried out QPO search. L.-X.D. led the theoretical modelling efforts and carried out calculations on disk formation. S.-X.W. and M.-J.L. performed X-ray spectral modelling in the IMBH scenario. L.L.T. simulated the spectra based on the reprocessing model. Z.-J.Z. analysed the outflow feature at 1\,keV. J.G. developed the XillverTDE model. Y.-L.W. and Q.-Y.W. are the duty TA of this peculiar transient. L.H., B.-F.L., F.-K., M.-J.L., Z.L., Y.-J.L., E.-L.Q., R.-F.S, R.S., R.S., B.Z. contribute to scientific explanation and provided comments on the manuscript.

S.-N.Z initially established the XIL at NAOC, CAS. C.Z., X.-J.S., S.L.S., X.-F.Z., Y.-H.Z., Z.-M.C. F.-S.C. and W.Y. contributed to the development of the WXT instrument. C.Z., Z.-X.L., H.-Q.C., D.-H.Z. and Y.L. contributed to the calibration of WXT data.
Y.L., H.-Q.C., C.-C.J., W.-D.Z., D.-Y.L., J.-W.H., H.-Y.L., H.S., H.-W.P. and M.-J.L. contributed to the development of WXT data analysis software.
Y.C., S.-M.J., W.-W.C., C.-K.L., D.-W.H., J.W., W.L., Y.-J.Y., Y.-S.W., H.-S.Z., J.G., J.Z., X.-F.Z., J.-J.X., J.M., L.-D.L., H.W., X.-T.Y., T.-X.C., J.H., Z.-J.Z., Z.-L.Z., M.-S.L., Y.-X.Z., D.-J.H., L.-M.S., F.-J.L., C.-Z.L., Q.-J.T. and H.-L.C. contributed to the development of the FXT instrument. 
S.-M.J., H.-S.Z., C.-K.L., J.Z. and J.G. contributed to the development of FXT data analysis software.

C.-C.J., L.-X.D., D.-Y.L., N.J., H.-Q.C., C.-W.Y., P.B., A.R., T.-G.W., H.-Y.Z., W.Y. contributed to the interpretation of the observations and the writing of the manuscript with contributions from all authors.

\paragraph*{Competing interests:}
The authors declare that they have no competing financial interests.

\paragraph*{Data and materials availability:}
Correspondence and requests for materials should be addressed to C.C.J.(ccjin@nao.cas.cn), L.X.D (lixindai@hku.hk), W.-M.Y. (wmy@nao.cas.cn), and C.Z.(chzhang@bao.ac.cn).
\clearpage

\subsection*{Supplementary materials}
Materials and Methods\\
Supplementary Text\\
Figs. S1 to S11\\
Tables S1 to S4\\
References \textit{(52-\arabic{enumiv})}\\ 


\newpage


\renewcommand{\thefigure}{S\arabic{figure}}
\renewcommand{\thetable}{S\arabic{table}}
\renewcommand{\theequation}{S\arabic{equation}}
\renewcommand{\thepage}{S\arabic{page}}
\setcounter{figure}{0}
\setcounter{table}{0}
\setcounter{equation}{0}
\setcounter{page}{1} 


\begin{center}
\section*{Supplementary Materials for\\ \scititle}

C.-C. Jin$^{\ast\dagger}$,
D.-Y. Li$^{\dagger}$,
N. Jiang$^{\dagger}$,
L.-X. Dai$^{\ast}$,
H.-Q. Cheng,
J.-Z. Zhu,
C.-W. Yang,
A. Rau,
P. Baldini,
T.-G. Wang
% ...
\\
\small$^\ast$Corresponding author. Email: ccjin@nao.cas.cn, lixindai@hku.hk, wmy@nao.cas.cn, chzhang@bao.ac.cn\\
\small$^\dagger$These authors contributed equally to this work.
\end{center}

\subsection*{This PDF file includes:}
Materials and Methods\\
Supplementary Text\\
Figures S1 to S11\\
Tables S1 to S4\\

\section{Materials and Methods}

\subsection{X-ray observations}

\subsubsection{EP-WXT observations} 
The Wide-field X-ray Telescope (WXT) is one of the two types of payloads on board the Einstein Probe (EP) mission \cite{Yuan2022, Yuan2025}. It employs the novel lobster-eye micro-pore optics (MPO) to enable a large instantaneous field-of-view (FoV) of 3850 deg$^2$ (combining all 12 WXT modules) and a sensitivity of $(2-3) \times10^{-11}$ $\rm erg\,cm^{-2}\,s^{-1}$ with 1\,ks exposure. EP240222a was first discovered by WXT during a calibration observation on 22 Feb. 2024, with a total exposure time of 21.9 ks. The detection significance is $>9$ with a net source counts of 88. The position of EP240222a was subsequently covered 12 times by WXT (see Table~\ref{tab:xray_label}). The WXT data reduction was performed following the standard data reduction procedures using the WXT Data Analysis Software (WXTDAS, Liu et al. in prep.).

\subsubsection{LEIA observations} 
The Lobster Eye Imager for Astronomy (LEIA) is a fully representative test model of EP-WXT \cite{Zhang2022, Ling2023}, with a FoV of 18.6 deg $\times$ 18.6 deg. It was launched on 27 July 2021 as an WXT pathfinder and has begun scientific observation since November 2022. The position of EP240222a was serendipitously covered by LEIA multiple times during its observations. An exposure time of 88 ks was obtained by merging 9 LEIA observations taken from 30 January to 9 February 2023, which did not detect any X-ray signal and provided an upper limit of $7.0 \times 10^{-13}$ erg s$^{-1}$ cm$^{-2}$ at 90\% confidence level for EP240222a at that time, assuming the same spectral shape with that of the first EP-WXT detection. Another 9 ks was obtained by merging 4 successive LEIA observations taken on 1 Feb. 2024, in which EP240222a was clearly detected with a significance of 6. Data reduction on the merged data was also performed with WXTDAS (Liu et al. in prep.).

\subsubsection{EP-FXT observations} 
The Follow-up X-ray Telescope [FXT; \cite{Chen2020}] is the other type of payload on board EP. Both FXT modules (FXT-A and FXT-B) adopt the classic Wolter-I type X-ray focusing optics. Multiple FXT Target-of-Opportunity (ToO) observations were triggered after the discovery of EP240222a. The first observation was carried out on 13 March 2024 with an exposure of 2 ks. Another observation with a longer exposure of 11.9 ks was triggered on 17 April 2024. The data were reduced using the FXT Data Analysis Software (FXTDAS\;v1.05) provided by the EP science center (EPSC) with the latest FXT calibration database (CALDB, v1.05)\footnote{http://epfxt.ihep.ac.cn/analysis}. The images taken by LEIA, EP-WXT, and EP-FXT are shown in Fig~.\ref{fig:xray_img}.

\subsubsection{Chandra observation} 
A Chandra Director's Discretionary Time (DDT) observation on EP240222a was triggered on 1 April 2024 using the High-Resolution Cameras (HRC-I), with an exposure time of 2 ks. Data reduction was performed following the standard procedures with the software Chandra Interactive Analysis of Observations [CIAO, v4.15; \cite{Fruscione2006}]. The data retrieved from the archive was reprocessed using the task {\it chandra\_repro} to apply the latest calibrations consistent with the current versions of CIAO and CALDB. The main objective of this observation was to obtain the best X-ray localization of EP240222a and confirm its association with the optical transient. The X-ray coordinates derived with the {\it wavedetect} task are RA $= $11$^h$32$^m$06$^s$.17, Dec $= +27^\circ00^\prime17^{\prime\prime}.6$ (J2000). The 1-$\sigma$ position error including both statistics and systematics is 0.73 arcsec.

\subsubsection{{\xmm} observation} 
To investigate the detailed X-ray spectral timing properties, an {\xmm} ToO observation was triggered on 23 May 2024, with an observation time of 54.6 ks. The data were reduced with the Science Analysis Software [SAS, v21.0.0; \cite{Gabriel2004}] along with the latest calibration files. The {\it evselect} tool was used to select good events from the three European Photon Imaging Cameras (EPIC, pn, MOS1 and MOS2) with negligible background flare contamination. Source spectra were extracted from a circular region of 35 arcsec, while the background spectra were extracted from a nearby source-free region. Then the {\it rmfgen} and {\it arfgen} tools were used to produce the response files and ancillary files. The {\it epatplot} tool was used to check and ensure that the spectra were not affected by the photon pile-up effect.

\subsubsection{NICER observations} 
Neutron star Interior Composition Explorer (NICER) is a scientific payload on board the International space Station (ISS) dedicated to the study of neutron stars. NICER began to monitor EP240222a from 14 March 2024, about one day after the first FXT observation. The monitoring campaign has a cadence of once per 2-3 days until the end of June. The data was downloaded from the HEASARC data archive center\footnote{https://heasarc.gsfc.nasa.gov/docs/archive.html}. For data reduction, the standard pipeline processing tool \textsc{nicerl2} was employed to generate the level 2 cleaned event files. The Complete Spectral Product Pipeline task \textsc{nicerl3-spect} was used to generate the spectral files and the Complete Light Curve Product Pipeline \textsc{nicerl3-lc} is used to generate the light curves of the source and background. The SCORPEON background model\footnote{https://heasarc.gsfc.nasa.gov/docs/nicer/analysis\_threads/scorpeon-overview} is utilized for the estimation of background. The exposure time for single observation is generally less than 2 ks due mainly to the interference of ISS structures. Furthermore, to increase the signal-to-noise, the data from March 2024 to June 2024 were combined by employing the {\it niobsmerge} tool. It should be noted that some of the NICER observations were carried out during when there was a strong geomagnetic activity with large Kp ($\ge5$) values reported. Strong non-X-ray flares contributed by high-energy particles were found in the data. To eliminate this effect, the good time intervals (GTIs) were generated by fine-tuning the threshold of the geomagnetic cutoff rigidity (COR-SAX). A careful visual inspection of the count rate within 0.3-2 keV (dominated by the source) was also performed to ensure the cleanness of the data.

\subsubsection{eROSITA observations} 
The extended ROentgen Survey with an Imaging Telescope Array instrument [eROSITA; \cite{Predehl21}] on board the Spectrum-Roentgen-Gamma [SRG; \cite{Sunyaev21}] observatory detected a source coincident with the position of EP240222a in each of its four eROSITA all-sky surveys (eRASS1-4) performed between 2019 and 2022 (Figure~\ref{eRASS_ima}). All spectra were extracted using the {\it srctool} tool of the eROSITA Science Analysis Software System [eSASS; \cite{Brunner22}] from event files version 020. We extracted source spectra with 40" radius circular regions, and background spectra following the procedure described in \cite{Liu22}. Spectral analysis was performed using the combined data from all seven telescope modules, both for the individual eRASS1-4 scans and for the combined eRASS:4 events. All the X-ray observation information and the derived unabsorbed fluxes are listed in Table~\ref{tab:xray_label}.

\subsection{X-ray spectral-timing properties}

\subsubsection{Pre-peak X-ray spectra} 
The analysis of the pre-peak eROSITA spectra was performed with the Bayesian X-ray Analysis software (BXA) version 4.1.2 \cite{Buchner14}, which connects the nested sampling algorithm UltraNest \cite{Buchner21} with the fitting environment CIAO/Sherpa \cite{Fruscione06}. The spectra were fitted unbinned and using C-statistic. The fitting procedure included a PCA-based background model [e.g., \cite{Simmonds18}] derived from a large sample of eROSITA background spectra \cite{Liu22}. For all spectra, three source models were fitted: a power-law, a multicolor disk model, or a blackbody, all modified by Galactic absorption [{\tt tbabs}; \cite{Wilms2000}]. All models can reproduce the eRASS:4 spectra (Figure~\ref{eRASS_spec}), with an inner disk temperature for the {\tt diskbb} model of T=$240^{+48}_{-36}$\,eV, a photon index of $\Gamma=2.62^{+0.31}_{-0.29}$ for the power-law model, and a blackbody temperature of T=$158^{+19}_{-17}$\,eV. Regarding the individual surveys, the eRASS1 spectrum is consistent with a non-detection, while for eRASS2-4, it is possible to constrain photon indices, inner disk temperatures and blackbody temperatures.

A comparison of the Bayesian evidence $Z$ suggests that the power-law model is overall preferred, while the {\tt diskbb} model is preferred over the black-body model. The derived unabsorbed fluxes are reported in Table~\ref{tab:xray_label}. 

\subsubsection{Peak X-ray spectra} 
To validate the spectral analysis, the spectra from LEIA and EP-WXT were binned with a minimum of 5 counters per bin given the small number of photons, and the C-statistic was used. The spectra from NICER, Chandra, EP-FXT and {\xmm} EPIC were binned to have 25 photons per group and $\chi^{2}$ statistic is adopted in the fit. All the X-ray spectral analysis described below was carried out using the {\sc Xspec} software [version 12.13.0c; \cite{Arnaud1996}].

The Galactic equivalent $N_H$ column density toward EP240222a is 1.8 $\times 10^{20}$\,cm$^{-2}$ \cite{willingale2013}, which is always taken into account and modelled by the {\tt tbabs} model with the abundance set to those in \cite{Wilms2000}. The intrinsic absorption to the source was modeled using {\tt ztbabs} with redshift $z$ fixed at 0.032. Firstly, we found that a multicolor disk model ({\tt diskbb}) modified by Galactic and intrinsic absorption ({\tt tbabs*ztbabs*diskbb}) provides good fits (reduced $\chi^{2} \sim 1.14$) to all the spectra below 1.7 keV. The best-fit inner disk temperature is $283_{-48}^{+66}$\,eV. An extra component is required above 1.7 keV, whose power law shape is clearly revealed by EPIC-pn with a photon index of 3.9$^{+0.2}_{-0.2}$.

However, such a steep hard X-ray power law will also extend into and dominate the X-ray flux below 0.5 keV, which is not physical because this component is generally believed to be produced by the Comptonisation of some hot electrons in the corona \cite{Pounds1995}, so there should be a low-energy cut-off. Therefore, we replace this phenomenological power law component with a more physical Comptonisation model {\tt compTT} \cite{Titarchuk1994}, whose seed photon temperature can be linked to the disk temperature of the disk component dominating the soft X-rays. This accretion disk plus corona model scenario provides good fits to all the X-ray spectra observed during and after the peak, since there was no significant spectral shape evolution. The best-fit disk temperature was 181$^{+15}_{-31}$\,eV, and the electron temperature of the corona was 586$^{+557}_{-159}$\,eV. The fitting results are summarized in Table~\ref{tab:xmm_fitting}. Figure~\ref{fig:xray_spec_xmm_nicer.png} shows two X-ray spectra and model fits. 

\subsubsection{Outflow feature at 1 keV} 
The outflow with the significant 1 keV feature is modeled with {\tt gsmooth*xillverTDE}. The {\tt xillverTDE} model \cite{Masterson2022} is a new flavor of the X-ray reflection model {\tt xillver} \cite{Garcia10}. To make it feasible for probing TDEs, the incident spectrum of {\tt xillverTDE} is a blackbody spectrum with $kT_{\rm bb} = 0.03-0.3\ {\rm keV}$. Recent research work shows that the reflected line profiles from super-Eddington systems are typically blue-shifted and broadened due to the relativistic outflow \cite{Thomsen19, Zhang24}. The {\tt xillverTDE} model includes a parameter, \textit{z}, to recover the velocity shift of the outflow. The predicted symmetrical broadening effect \cite{Thomsen19} is accounted for a {\tt gsmooth} model.

The $kT_{\rm bb}$ of {\tt xillverTDE} and $kT_0$ of {\tt comptt} is linked to the $T_{\rm in}$ of {\tt diskbb}, and the new result is $T_{\rm in}=0.18\pm0.03\ {\rm keV}$. The metallicity $A_{\rm Fe}$ and inclination $i$ are fixed to 1 and $45^\circ$ separately. The gas density pegged then is fixed at $n_e=10^{19}\ {\rm cm^3}$. These procedures were also adapted by previous works \cite{Masterson2022,Yao2024}. The best-fit $z=-0.30\pm 0.03$ corresponds a outflow velocity $v_{\rm out} = 0.34\pm0.03c$. This result indicates the 1 keV feature is from the blueshifted Oxygen K-shell emission. The best-fit gaussian $\sigma$ is $0.08\pm0.06\ {\rm keV}$. The plasma temperature of the {\tt comptt} is $kT = 0.28\pm0.07\ {\rm keV}$, and the plasma optical depth $(\tau_p)$ is pegged at 200 then fixed. The column density of the partial absorption component is $N_h=0.1\pm0.5\times10^{22}\ {\rm cm^{-2}}$, the ionization parameter is $\log \xi=-2.2\pm136.3$ and the covering fraction is $0.1\pm 0.3$.

\subsubsection{Non-detection of X-ray QPO} 
The individual observations of NICER/XTI and EP-FXT was split into several GTIs, resulting in several continuous data segments. We first divided the NICER data into 500 s continuous segments and extracted their light curves with a time resolution of 1 s. This resulted in a total of 12 light curve segments, with a total exposure time of 6 ks. Considering that the mean exposure time is mostly less than 550 seconds per segment, choosing 500 seconds ensures that the power density spectrum (PDS) can be extended to a lower frequency limit of $2\times10^{-3}$ Hz. A Leahy normalized (Poisson noise level of 2) PDS was extracted from each of these individual light curves and they were all combined to obtain the average PDS in Figure~\ref{fig-pds}a. For the FXT observations, 5 segments of 1.8 ks each was investigated, which resulted in the average PDS in Figure~\ref{fig-pds}b. Both of the PDS lead to the same conclusion that no significant QPO signal above 0.005 Hz, and they are dominated by red noise at low frequency.

\subsection{X-ray Spectral Modelling in the IMBH scenario}

We examined the disk parameters by fitting the spectra from the three EP-FXT epochs (epoch-1 to 4) and one {\xmm} epoch (Table~\ref{tab:Xrayspfit}). The X-ray emission of epoch-1 aligns with the disk emission, whereas the other epochs exhibit a hard component that necessitates an additional source to account for it. We constrain the disk parameters by fitting the spectra using a slim disk model \cite{Wen2021,Wen2022}. 
Similar to the cases of XMMJ2150 \cite{Wen2021} and 3XMM J150052.0+015452 \cite{Cao2023}, the brightness of the disk is insufficient to adequately fit the spectrum in epoch-1 under the original spectrum hardening assumptions. This result suggests that the source experienced highly super-Eddington accretion during epoch-1, which is consistent with theoretical predictions for main-sequence IMBH disruptions [e.g. \cite{Rees1988, Chen2018}]. For high super-Eddington accretion rates, the X-ray luminosity of the slim disk is largely insensitive to the exact super-Eddington accretion rate value but does depend on the choice of spectrum hardening factor $f_{\rm c}$ \cite{Wen2020}. In this analysis, we treat $f_{\rm c}$ as a free parameter with a flat prior ranging between 2.0 and 3.0. 

We fit the epoch-1 spectrum separately using an absorbed slim disk model \cite{Wen2021}, while the other spectra were fitted simultaneously with an absorbed slim disk model [{\tt slimd}); \cite{Wen2022}] plus a thermal Componization model [{\tt THcomp}; \cite{Zdziarski_2020}]. During the fitting, we used the same black hole mass ($M_\bullet$), black hole spin ($a_\bullet$), and disk inclination angle ($\theta$) for all epochs, but allowed the accretion rate ($\dot m$), absorption parameter ($N_H$), optical depth ($\tau$), and corona temperature ($\kappa T_e$) to vary. Our analysis shows that tying the $N_H$, $\tau$, and $\kappa T_e$ parameters across all EP-FXT spectra only reduced the $Cstat$ by 2.1, while the number of free parameters decreased by 6. This suggests that the absorption environment and the shape of the hard component remained stable, and so in the simultaneous fitting, we used the same $N_H$, $\tau$, and $\kappa T_e$ for all the epochs. Since the {\xmm} epoch was observed on the same day as epoch-3, we set the same disk parameters for the two spectra, but allowed the absorption and optical depth to vary to account for possible discrepancies due to the different energy bands of the two telescopes.

The fitting results are listed in  Table \ref{tab:Xrayspfit}. We achieved a good fit for epoch-1 with $Cstat/\nu = 0.91$, suggesting that the spectrum is consistent with a disk spectrum. The black hole mass $M_\bullet$ and spin $a_\bullet$ were constrained to $5^{+3}_{-1} \times 10^4 M_\odot$ and $>0.9$, respectively. The accretion rate $\dot m$ was constrained to be $\dot m > 7$, aligning with the assumption of highly super-Eddington accretion. For the simultaneous fits, we obtained a good fit with a total $Cstat/dof = 1407.5/1333 = 1.06$. The mass $M_\bullet$ and spin $a_\bullet$ were constrained to $7.7 \pm 4 \times 10^4 M_\odot$ and $>0.4$, consistent with the results from epoch-1. We also fitted the spectra by replacing the slim disk model with the {\tt tdediscspec} model \cite{Mummery2021, Mummery2023}. The results are also listed in  Table \ref{tab:Xrayspfit}. We obtained a good fit for the epoch-1 spectrum with $Cstat/\nu = 0.91$ and a good fit for the simultaneous fit epochs with a total $Cstat/dof = 1412.8/1332 = 1.06$. The disk peak temperature $T_p$ decreased from $1.73 \times 10^6$ K to $1.64 \times 10^6$ K over two months, indicating a possible drop in accretion.

Figure \ref{ma} shows the constraints on $M_\bullet$ and $a_\bullet$ from the simultaneous fits. The contours represent the results from the {\tt slimd} fitting. The $M_\bullet$ from the {\tt tdediscspec} model is estimated by $M_\bullet = (4.9 \pm 4.5) \times 10^4 \times R_p/(10^{10} \text{cm})$\cite{Mummery2023}. We constrain the $M_\bullet$ to be $8\pm4\times10^4M_\odot$ with a slim disk model. Due to the degeneracy between the $M_\bullet$ and $a_\bullet$, the $a_\bullet$ is constraint to be $>0.7$. The $M_\bullet$ constrained here is consistent with the $M_\bullet$ derived from FXT-EP 1, as listed in  Table \ref{tab:Xrayspfit}. These results strongly support the conclusion that this TDE is associated with an IMBH.

\subsection{Optical Photometric observations}

\subsubsection{Xinglong Schmidt 60/90cm Telescope}
On 12 March 2024, a follow-up campaign was triggered using the Schmidt Telescope (60/90 cm) at Xinglong Observatory in Hebei, China, with observations conducted in the {\it clear} band. The images were stacked using {\tt SWarp}\cite{Bertin2010}, and PSF photometry was performed on the stacked images using AutoPhOT (https://github.com/Astro-Sean/autophot/)\cite{AUTOPHOT}. An optical transient at RA = $11^h32^m06^s$.1, DEC = $27^\circ00^\prime18^{\prime\prime}$ with a magnitude of $r$=21.2$\pm$0.2 was detected within the error circle of the FXT detection. The photometric result was calibrated using $r$-band PS1\cite{Flewelling2020} standard stars.

\subsubsection{Xinglong 2.16m Telescope}
Xinglong 2.16m optical telescope \cite{Fan2016} was triggered via a ToO observation on 13 March 2024. The photometric observations were carried out with the white-light (no filter) band attached to the Beĳing Faint Object Spectrographa and Camera (BFOSC). The source were monitored over the following days on 14 and 19 March. The observation information is summarized in  Table~\ref{table:catalog_opt}. After the bias and flat field correction, the images were stacked by the {\it imcombine} tool in {\sc IRAF}. The photometric measurements were carried out using the Source-Extractor tool {\it SExtractor}. The automatic aperture photometry was derived from Kron's ``first moment" algorithm \cite{Kron1980}. Then the instrumental magnitude was calibrated to Pan-STARRs DR1. 
The optical counterpart can be clearly detected in the images at the position RA = $11^h32^m06^s$, Dec = $27^\circ00^\prime18^{\prime\prime}$.

\subsubsection{WFST}
The Wide Field Survey Telescope (WFST) is a dedicated photometric survey facility equipped with a 2.5-meter diameter primary mirror and a 0.73 gigapixel mosaic CCD camera with an effective field of view of 6 square degrees~\cite{WangT2023}. The WFST achieved its first light in September 2023 and began the pilot survey in March 2024 after a six-month commissioning period. 
We initiated a follow-up campaign with WFST in the $ugr$ bands with three exposures of 120s each day from UT 14 March 2024 to 7 July 2024. 

We stacked the images in the same band every day using the {\tt SWarp}~\cite{Bertin2010} and then performed PSF photometry on the stacked images using the {\tt Photutils} package of Astropy for the WFST data. The g- and r-band photometric results were calibrated using PS1~\cite{Flewelling2020} standards in the field of view, while the $u$-band was calibrated using SDSS standards. All photometric data were corrected for the Galactic extinction of E(B-V)=0.017 mag~\cite{SF2011}. An optical transient (RA = $11^h32^m06^s$.15, DEC = $27^\circ00^\prime18^{\prime\prime}.1$) with magnitudes of $u$=21.33$\pm$0.25, $g$=21.28$\pm$0.08, and $r$=21.55$\pm$0.17 was identified within the error circle of the X-ray transient on 19 March 2024. No significant variability was detected in the following weeks. Compared to the faint source detected at the same position in the DR10 archive  ($g$=23.98, $r$=23.43), the source has brightened by about 2 magnitudes. Information of the optical observation taken by Xinglong telescopes and WFST is summarized in Table~\ref{table:catalog_opt}

\subsubsection{GROND}
EP240222a was observed with the Gamma-ray Burst Optical Near-infrared Detector [GROND; \cite{Greiner2008}] mounted at the MPG 2.2m telescope at ESO's La Silla observatory on 14 March 2024 at 04:11 UT. Observations were performed simultaneously in the J- and H-bands with an exposure of 15\,min and at a mean airmass of 1.8. The data were reduced using the standard IRAF-based GROND pipeline \cite{kruehler08}. The aperture photometry was calibrated against the Two Micron All Sky Survey catalogue [2MASS; \cite{skrutskie06}] and converted into the AB system. No source was found at the position of the WFST and Chandra position to 3$\sigma$ limiting AB magnitudes of $J$$>$20.4 and $H$$>$19.7.

\subsubsection{ZTF} 
EP240222a is located in the field of the Zwicky Transient Survey (ZTF), providing the opportunity to check the historical light curves. We have retrieved all publicly available photometric data for EP240222a from the ZTF forced-photometry service~\cite{Masci2023}. These data were obtained using PSF photometry on the difference images, with the position fixed to the WFST coordinate. A notable optical flare was identified starting from November 2023, coinciding with the X-ray flare. This result suggests that the optical flare has persisted for several months and its color ($g-r=-0.2$) appears to be consistent with optical TDEs~\cite{vV2021}.

\subsection{Optical Spectroscopic Observations}

We took the optical spectra of EP240222a with the Optical System for Imaging and low-Intermediate-Resolution Integrated Spectroscopy [OSIRIS; \cite{Cepa2003}] mounted on the Gran Telescopio Canarias (GTC; 10.4m telescope) on 20 March 2024. 
We used the R1000B grating with peak efficiency at a wavelength of 5455 \AA, covering a range of approximately 3630 to 7500 \AA, and achieving a resolution of $\sim$ 1000. The observations had a total exposure time of 3 hours. A total of six images were taken, each over a half-hour. Concurrently, we observed the spectrum of star Ross640 for flux calibration and HgAr and Ne lamp for wavelength calibration. We reduced the raw spectra using standard IRAF software. The median signal-to-noise ratio of the spectra is 4.5. Due to the overall low signal-to-noise ratio of the GTC spectrum, and the simultaneous photometry provided by WFST during this phase of the outburst, we recalibrated the spectrum using the photometric values given by WFST around March 20, specifically $u$ = 21.32 mag, $g$ = 21.34 mag, and $r$ = 21.57 mag.

The mass of the stellar system of EP240222a is small, and the stellar population model fit to the optical photometric data indicates a low internal extinction before the outburst (see Figure~\ref{fig:hostfit}). For the above reasons, and due to the lack of clear extinction features in the spectrum, we performed only a Galactic extinction correction on the spectra, using correction values the same as the photometry data [E(B-V) = 0.017; \cite{SF2011}]. Since this correction value is small, the spectra before and after correction show only slight differences at the blue end. Subsequent line and spectra shape measurements based on the optical spectra were performed using the spectra corrected for Galactic extinction. The spectrum shape exhibits a blue feature without distinct stellar features. Figure~\ref{fig:gtcspec} shows the results of the optical emission line fittings. Two emission lines are clearly visible at observed wavelengths of 4844 \AA\ and 6773 \AA: we identify them as He {\sc II} $\lambda$4686 and H$\alpha$, yielding a redshift of 0.032. Using a single Gaussian model to fit these two lines, we obtained FWHMs of $930\pm220\mathrm{km\ s}^{-1}$ and $1340\pm170\mathrm{km\ s}^{-1}$, respectively, with fluxes of $6.2\times10^{-17}\mathrm{erg\ s^{-1}\ cm^{-2}}$ and $9.5\times10^{-17}\mathrm{erg\ s^{-1}\ cm^{-2}}$. At a redshift of $z=0.032$, their luminosities are $1.5\times10^{38}\mathrm{erg\ s}^{-1}$ and $2.2\times10^{38}\mathrm{erg\ s}^{-1}$, respectively.

EP240222a was also observed on 2 May 2024, using the Gemini Multi-Object Spectrograph [GMOS; \cite{Hook2004}] mounted on the Gemini North telescope, under the program GN-2024A-DD-103. The observations were conducted using the B480+\_5309 grating and  a slit width of 1.0 arcsec. A total of 12 exposures of 900s  were obtained,  with four exposures centered at 5000 \AA, four at 5100 \AA, and four at 5200 \AA, to compensate for the detector gaps. Data were processed with {\tt Pypeit}~\cite{Pypeit1,Pypeit2}, a highly automated tool for spectroscopic data reduction.  Flux calibration was performed using the standard star BD+28D4211, which was observed with a similar configuration but only at a central wavelength of 5100 \AA\ on a different night. The resulting spectrum has a median signal-to-noise ratio of 20, which is about 5 times higher than that of GTC spectrum. We robustly detected the emission lines of $\mathrm{H}\alpha$, $\mathrm{H}\beta$, and high-order Balmer lines, as well as $\mathrm{He\,\textsc{ii}}$ $\lambda 4686$\ and $\mathrm{N\,\textsc{iii}}$ $\lambda 4640$. These lines were modeled using a single Gaussian function and a linear function for the local continuum, resulting in flux values of $\rm 9.6\pm0.2\, \mathrm{erg\,s^{-1}\,cm^{-2}}$, $\rm 2.1\pm0.3\, \mathrm{erg\,s^{-1}\,cm^{-2}}$, $\rm 10.2\pm0.4\, \mathrm{erg\,s^{-1}\,cm^{-2}}$ and $\rm 5.8\pm0.7\, \mathrm{erg\,s^{-1}\,cm^{-2}}$, with corresponding luminosities of $2.3\times10^{38}\mathrm{erg\ s}^{-1}$, $0.5\times10^{38}\mathrm{erg\ s}^{-1}$, $2.4\times10^{38}\mathrm{erg\ s}^{-1}$, and $1.4\times10^{38}\mathrm{erg\ s}^{-1}$,  as well as FWHM values of $\rm 1141\pm32\,km\,s^{-1}$, $\rm 735\pm123\,km\,s^{-1}$, $\rm 1008\pm47\,km\,s^{-1}$ and $\rm 1446\pm160\,km\,s^{-1}$ for $\mathrm{H}\alpha$, $\mathrm{H}\beta$, $\mathrm{He\,\textsc{ii}}$ and $\mathrm{N\,\textsc{iii}}$ respectively. With the line centers of $\mathrm{H}\alpha$ and $\mathrm{H}\beta$ obtained from the fitting process, we calculated the redshift to be $0.03251 \pm 0.00013$, which matches with that of the nearby large galaxy within errors. 

\subsection{Radio Observations\\}
We observed EP240222a using VLA on 4 and 7 April 2024, under the DDT program with code 24A-457. The observations were first performed at the Ku-band and then C-band, centered at 15 GHz and 6 GHz, respectively. To solve for the time-dependent complex gains, we used the nearby phase calibration source J1125+2610, while the standard calibrator 3C 147 was used as bandpass calibrator and to set flux density scale. The data were reduced following standard procedures with the CASA package \cite{CASA2022}. According to the pipeline log files, we examined each spectral window and flagged abnormal data due to RFI or hardware issues. The calibrated data were then selected and we used the {\tt CLEAN} algorithm to remove possible contamination from side-lobes, with the conventional Briggs weighting and ROBUST parameter of 0. The final cleaned maps in both bands have a similar rms noise of $\sim10\mu$Jy/beam, measured using the {\tt IMFIT} task in CASA. We do not detect EP240222a in either C-band or Ku-band, resulting in a 5$\sigma$ upper limit of $\sim50\mu$Jy. In the field of view of VLA imaging, another radio source about 53$^{\prime\prime}$ away from EP240222a is clearly detected at both bands. We used the {\tt IMFIT} task in CASA to fit the radio emission component with a two-dimensional elliptical Gaussian model to determine the position, the integrated and peak flux density. The radio source's position is spatially coincident with the nearby galaxy, 2MASX J11320214+2700207. The integrated flux is 245$\pm$27 $\mu$Jy (including the flux calibration uncertainty that is assumed to be of 5\% of the flux density) and 206$\pm$12 $\mu$Jy at the C-band and the Ku-band, respectively, suggesting a radio spectral index of $-0.19$ ($S_{\nu}\propto \nu^{\alpha}$ ).

\subsection{Broadband SED properties\\}

 Figure~\ref{fig:xray_opt_sed} shows the broadband spectral energy distribution (SED) of EP240222a, along with those of some typical thermal TDEs. The X-ray spectrum of EP240222a is from the first FXT observation, and the three optical data points are from WFST. EP240222a is distinct in that its X-ray luminosity is two orders of magnitude higher than its optical luminosity, marking it as a rare X-ray bright TDE. The bolometric luminosity of EP240222a is dominated by its X-ray luminosity. Considering the source may be in a super-Eddington accretion state, we used the {\tt agnslim} model \cite{Kubota2019} in {\sc xspec} to fit the multi-wavelength SED data. This model assumes that the optical radiation originates from the accretion disk. It yielded a minimal chi-squared of 36.5 for 30 degrees of freedom, suggesting a reasonably good fit. The best-fit black hole mass is $5.3^{+5.0}_{-2.2}\times10^{4}$ M$_{\odot}$,  and the mass accretion rate is $2.74^{+16.38}_{-1.40}$. If we consider that part of the optical radiation is due to X-ray reprocessing, 
the contribution of the accretion disk to the optical would be smaller. Then to conserve energy across all wavelengths and produce the observed X-ray luminosity, a smaller black hole mass and a higher spin would be required. Consequently, the special SED of EP240222a supports the presence of an IMBH.

\subsection{Properties of the host\\}

The host of EP20240222a was detected in the $g$, $r$, $i$, and $z$ bands with fluxes of 0.927$\pm$0.106, 1.538$\pm$0.171, 1.952$\pm$0.199 and 3.014$\pm$0.426 $\mu$Jy in DECaLS \cite{Dey2019}. To infer the physical properties of the host from these photometric data and the upper limit in $y$-band of the Pan-STARRS survey, we employed the public Bayesian SED modeling code iSEDFIT \cite{Moustakas2013,Moustakas2017}. iSEDFIT generated a Monte-Carlo grid of galaxy models encompassing various star formation histories, stellar metallicity, using the BC03 stellar population synthesis library, and dust content. We analyzed 25,000 galaxy models that span ages from 0.1 to 13 Gyr, bursting duration $\tau=0.1-5$ Gyr, metallicity $Z=0.004-0.03$, and extinction $A_v=0.1-2.0$, and using the Chibier IMF. The results are shown in the  Figure \ref{fig:hostfit}. The stellar mass was determined to be in the range of $\log M_* (M_\odot)= 6.52-7.28$ (90\% probability) with a median value of 6.94. We have also tried other SED fitting procedures, which all give a similar stellar mass, i.e., the fit of CIGALE~\cite{CIGALE} gives a  $\log M_* (M_\odot)= 7.02\pm0.32$. Using the scaling relation between the black hole and the bulge stellar mass \cite{Schutte2019}, we estimate a black mass of order $10^4$ $M_\odot$. Using the relationship for early-type galaxies \cite{Greene2020}, a similar mass of $10^{3-4}$ $M_\odot$ can be obtained. However, the intrinsic scatter of these scaling relations at the lower mass end is large.

\subsection{The nearby large galaxy\\}
The centre of the nearby large galaxy in the same redshift ($z=0.03275$), 2MASX J11320214+270020, is at a projected distance of 53.1$^{\prime\prime}$ from EP240222a. It was spectroscopically observed by the SDSS in 2005, which suggests a typical old galaxy dominated by absorption lines. This is consistent with its elliptical morphology.  The stellar mass and star formation rate inferred from SED fitting to optical and infrared photometry are $7.8\times10^{10}$~$M_\odot$ and $5.0\times10^{-3}$~$M_\odot~\rm yr^{-1}$~\cite{Chang2015}, respectively. The central black hole mass estimated from the empirical correlation with the host bulge mass is $3.6\times10^{8}$~$M_\odot$~\cite{KH2013}. 
2MASX~J11320214+270020 is the central galaxy of a dark matter halo with a mass of about $3.5\times10^{12}~M_\odot~h^{-1}$ according to the halo-based group finder~\cite{Yang2007}.
Interestingly, it is detected in the radio with a relatively flat spectral index of $-0.19$ ($S_{\nu}\propto \nu^{\alpha}$ ). The galaxy remains compact at the radio band at the resolution of $\sim1^{\prime\prime}$. The radio luminosity at 6 GHz is $\sim5.8\times10^{20}$ W Hz$^{-1}$, indicating that the galaxy is formally radio quiet. The stellar shells in the outer region of 2MASX\,J1132+2700 indicate that it might have experienced minor mergers with the satellite galaxies around \cite{Cooper2011}.

\subsection{Disk formation timescale\\}
Simulations have shown that one of the most promising mechanisms for forming TDE disks is through the collision of the debris stream with itself due to general relativistic apsidal precession \cite{Shiokawa15, Bonnerot2016}. In this picture, the debris orbital energy dissipates through stream collisions, the power of which depends on the black hole mass $M_{\rm BH}$, stellar mass $m_\star$ and orbital penetration parameter $\beta$. First-order calculations \cite{Dai15, Wong22} gives that the disk formation efficiency can be defined as:
\begin{equation}
    \mathcal{C}\left(M_{\rm BH},\beta,m_\star\right) \equiv \frac{\Delta E_{\text{collision}}}{\Delta E_{\text{total}}} , \label{eq:circeff}
\end{equation}
where $\Delta E_{\text{collision}}$ is the specific energy loss at the stream self-crossing point, and $\Delta E_{\text{total}}$ is the total energy loss needed for the debris for fully circularize. If $\mathcal{C}\sim1$, it means that a circularized debris disk can form within one fallback timescale $T_{\rm fb}$. The actual disk formation/circularization timescale should be much longer if $\mathcal{C}\ll 1$.
We further assume that the time that the disk is fully assembled and circularized can be approximated by 
\begin{equation}
    T_{\rm circ} = T_{\rm fb} / \mathcal{C} \label{eq:Tcirc}.
\end{equation}
For a solar-type star, the disk circularization timescale depends on $M_{\rm BH}$ and $\beta$, which is plotted as in  Figure~\ref{fig:Tcirc}.  

The disk formation time of EP240222a can be constrained using the eROSITA pre-peak and X-ray/optical post-peak observations. Assuming that the TDE onset is around the first eROSITA detection, and the TDE flare peaks in fall 2023, we estimate the disk circularization time to be 1200 days. Furthermore, stars have higher chances to be disrupted along low-$\beta$ orbits. Using $\beta\sim 1$, we show in  Figure~\ref{fig:Tcirc} that EP240222a is most likely from a black hole of  $(1-4)\times10^5 M_\odot$. However, the start time of the rising phase can be much earlier than the first eROSITA detection, so the black hole mass estimated above is only an upper limit, and so it is also consistent with EP240222a being an IMBH-TDE.

\subsection{Origin of optical emissions during the plateau phase\\}
We employ the sophisticated Monte Carlo radiative transfer code, Sedona, to model the optical emissions observed in EP240222a using a reprocessing model. Sedona accounts for Comptonisation effects and tracks various emission and absorption processes, including free-free, bound-free, and bound-bound, in the scattering-dominated regime \cite{Kasen06, Roth16}. 
A key feature of Sedona is the determination of the excitation and ionization state of the matter by solving the non-local thermodynamic equilibrium (non-LTE) equations under statistical equilibrium. This allows us to account for deviations from thermal equilibrium that can significantly impact the ionization and excitation populations and hence the emission spectrum.

A 1D spherical geometry is used to represent the reprocessing envelope, with the density varying as a power-law function of the radius: $\rho(r) = \rho_0 (r/r_0)^{-q}$. The inner boundary is set to $r_0=10R_g$, where the gravitational radius is defined as $R_g=GM_{\rm BH}/c^2=1.5\times 10^{10} {\rm \ cm} \ (M_{\rm BH}/10^5 M_\odot)$.
The gas density at the inner boundary, $\rho_0$, is determined by constraining the integrated mass from $r_0$ to the outer boundary of the envelope (set at $r_{\rm max}=6000R_g$) to be the envelope mass $M_{\rm env}$. The gas outflow velocity is chosen such that the outflow mass rate $\dot{M}_{\rm out} = 4 \pi r^2 \rho(r) v(r)$ is conserved. This gives the dependence of the outflow velocity $v(r) = v_0 (r/r_0)^{q-2}$, where $v_0$ is the outflow launching velocity at $r_0$. Finally, we inject a blackbody spectrum at the observed X-ray temperature $T_X \sim 3.15\times10^6 K$ from the inner boundary of the gas envelope. This process is repeated for at least 20 iterations until the escaped spectrum, ionization, and gas temperature reach equilibrium.


Approximately 1,000 simulations with varying parameters were carried out and the escaped spectra were compared to the observed TDE optical emission spectrum as observed by Gemini on 2 May 2024. Specifically, the free parameters and their ranges are: black hole mass $M_{\rm BH} \in [5\times10^4, 10^6] M_\odot$, gas envelope mass $M_{\rm env} \in [10^{-6}, 0.1] M_\odot$, gas density slope $q \in [0.25, 2]$, and injected X-ray luminosity $L_X \in [0.5\times10^{43}, 1.5\times 10^{43}] \ \mathrm{erg \ s^{-1}}$. Also, we adopt two different gas outflow launching velocities $v_0=0.01c$ and $0.1c$. To determine the best fit between our modelled spectrum and the Gemini optical spectrum (after excluding emission line regions), we compare the reduced Chi-square statistic, defined as $\chi^2_\nu= 1/\mathrm{d.o.f.} \sum (y_{\rm obs}-y_{\rm model})^2/ \sigma_{\rm y, \ obs}^2$, where $\mathrm{d.o.f.}$ represents the degrees of freedom.

The best-fit model shown in  Figure~\ref{fig:Best_fit_SED} is obtained with the parameters $M_{\rm BH}=10^5 M_\odot$, $M_{\rm env} = 0.002 M_\odot$, $q = 0.5$,  $L_X =1.5 \times 10^{43} \rm erg \ s^{-1}$, $v_0=0.01c$, and $\chi_\nu = 70.7 /d.o.f. = 1.7$. 
In this figure, we also show the spectral observations of this TDE for comparison. The upper panel shows that the X-ray emission injected into the envelope (black dotted curve) has a temperature similar to the observed X-ray emission (purple line) and slightly lower luminosity. The injected X-ray emission is reprocessed inside the envelope into UV and optical emissions through similar processes as discussed in \cite{Dai18, Thomsen22}. The reprocessed emission spectrum (black solid line) is compared with the Gemini spectrum (red solid line). In the lower panel, we zoom into the optical bands and show the comparison between model and observation. It can be seen that our model can recover the observed SED well. 

Our modelling results of the optical emissions seen in EP240222a further support that: 1) This TDE occurs around an IMBH with $M_{\rm BH}\sim10^5 M_\odot$. 2) Fast outflows are launched from this event. 3) Reprocessing of X-ray photons in such fast outflows can be responsible for the origin of the optical emissions as observed.
Furthermore, the 1D nature of our model means that we cannot simultaneously model the X-ray and optical emission. In a more physical scenario, such as when a super-Eddington accretion flow is formed around the black hole, an observer viewing the accretion flow from a face-on orientation will see the X-ray emissions escaped from the funnel close to the pole and the reprocessed optical emissions produced from fast outflows at the same time.

\subsection{Event rate estimation\\}
EP240222a is the first TDE detected by EP after its successful launch in 9 Jan. 2024, with the first EP-WXT detection on 22 Feb. 2024. It is currently the only IMBH-TDE candidate discovered by EP. EP240222a shares many similarities with XMMJ2150, both located in the outskirts of galaxies, but EP240222a is much better observed, including the first spectroscopy observation of this type of transients, making it a representative of its rare population. For EP
240222a-like TDEs, assuming a typical peak luminosity of $10^{43}$ erg s$^{-1}$ and a plateau phase duration of one year, we can calculate the effective monitoring volume based on WXT's field-of-view and sensitivity curve \cite{Yuan2022}. From 19 Jan. 2024 to 1 June 2024, WXT conducted over 2420 observations, covering a total monitored volume of $2.26 \times 10^8$ Mpc$^3$. Therefore, the event rate is estimated to be $4.4 \times 10^{-9}$ Mpc$^{-3}$ yr$^{-1}$. 
This rate is consistent the recent results by \cite{Chang24}, who have calculated IMBH-TDE rates based on the realistic stellar profiles of galaxies or clusters hosting IMBH candidates and obtained a volumetric TDE rate of $\sim 10^{-9}-10^{-8}$ Mpc$^{-3}$ yr$^{-1}$ for IMBHs. Since the host type of EP240222a is unclear, its number density is unknown. However, we can adopt a number density of $1.4 \times 10^{-2}$ Mpc$^{-3}$ for normal galaxies like 2MASX J11320214+2700207 in the local Universe \cite{Donley2002}, then the corresponding event rate is $3.2 \times 10^{-7}$ gal$^{-1}$ yr$^{-1}$. These results are also consistent with the event rate estimated by \cite{Lin2018} based on the discovery of XMMJ2150 from the {\xmm} data archive. Furthermore, the detection of EP240222a by EP soon after its launch is partly due to WXT operating in 0.5-4 keV, making it more sensitive to the higher thermal temperatures of such TDEs. EP's observations in the coming years are expected to detect more similar cases in their peak outburst phase, providing better constraints on the occurrence rate of this rare TDE population.

\newpage




\setcounter{figure}{0}
\setcounter{table}{0}

\begin{figure}[h]%
\centering
\includegraphics[width=\textwidth]{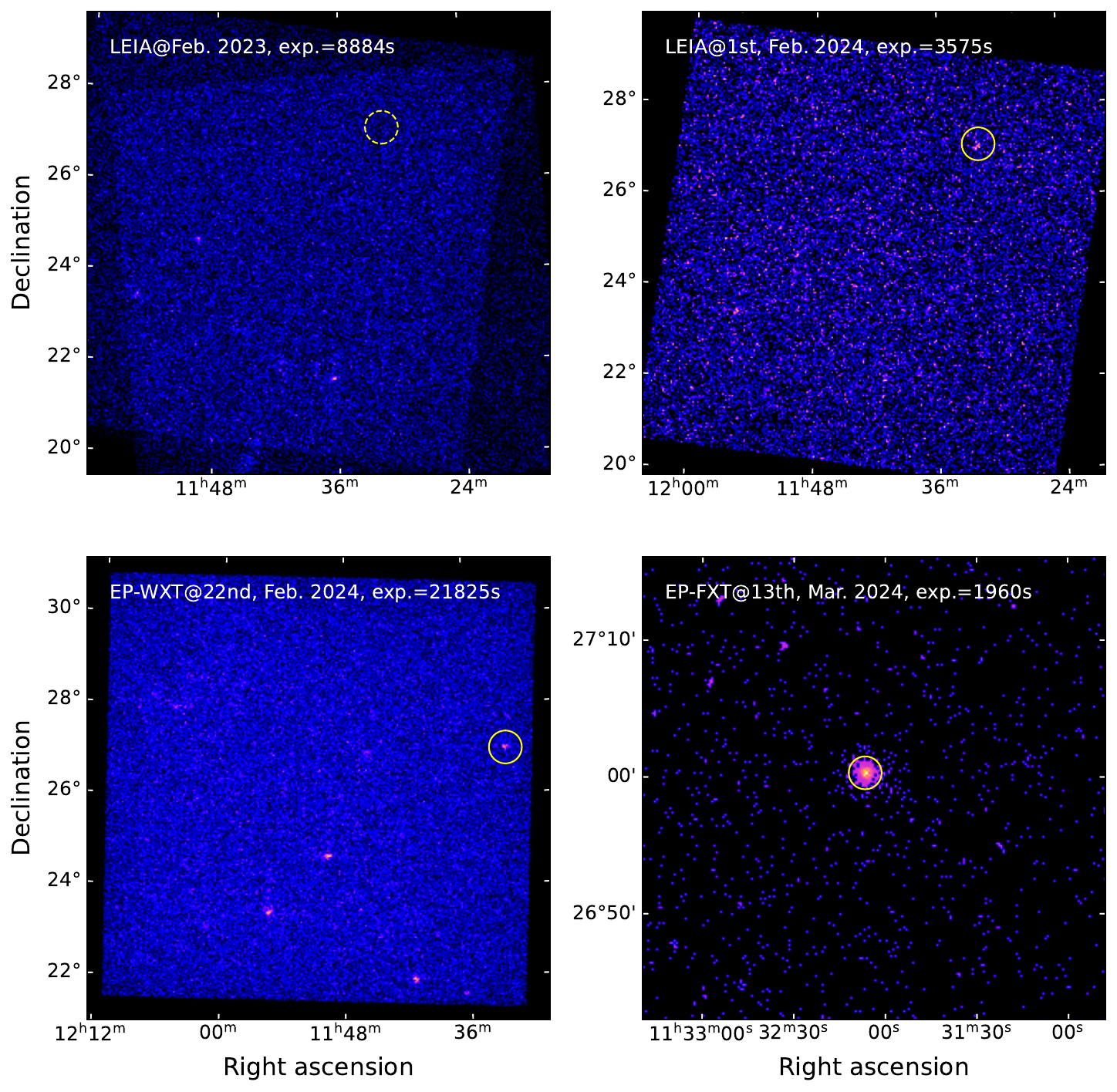}
\caption{Soft X-ray images of EP240222a taken by LEIA in Feb. 2023 (top left), Feb. 2024 (top right), EP-WXT in Feb. 2024 (bottom left), and EP-FXT in Mar. 2024 (bottom right). The localizations of EP240222a are indicated by yellow circles in each panel. }
\label{fig:xray_img}
\end{figure}

\begin{figure}[h]%
\centering
\includegraphics[width=0.5\textwidth]{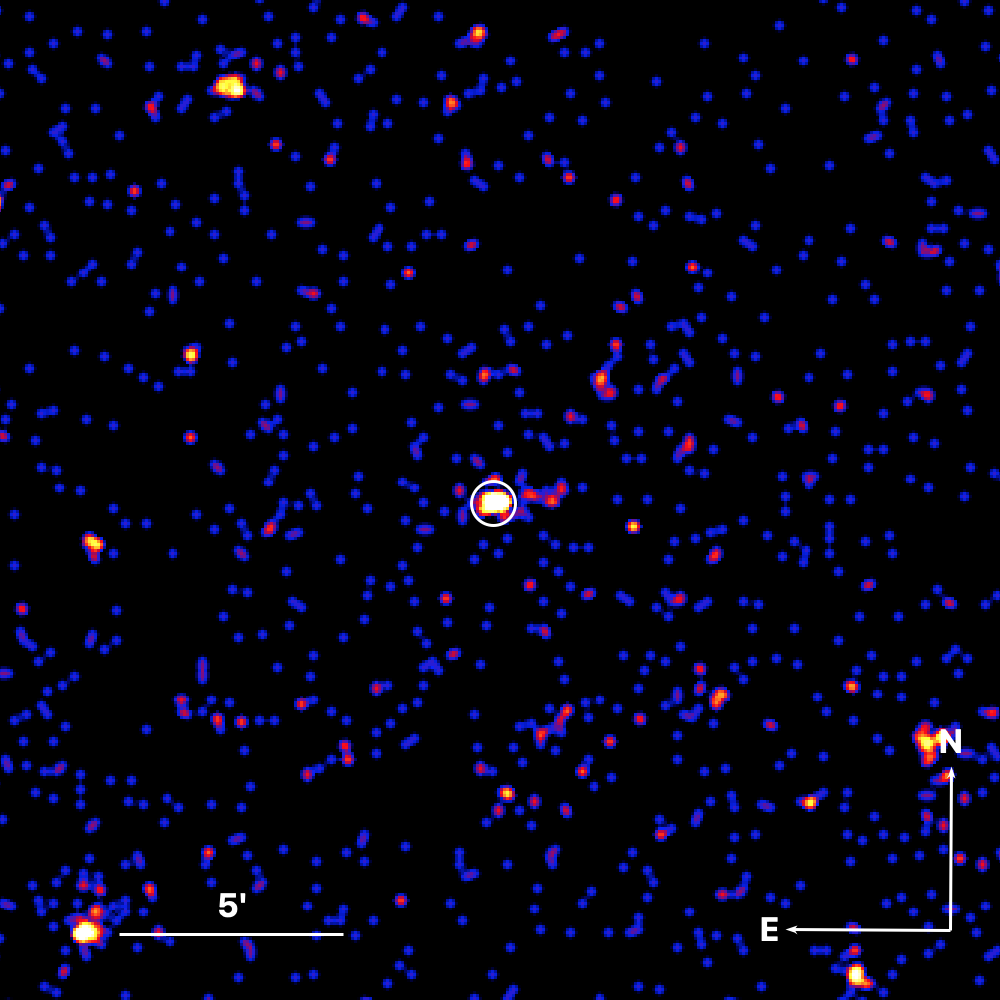}
\caption{eRASS:4 combined image centered on EP240222}\label{eRASS_ima}
\end{figure}

\begin{figure}[h]%
\centering
\includegraphics[width=1\textwidth]{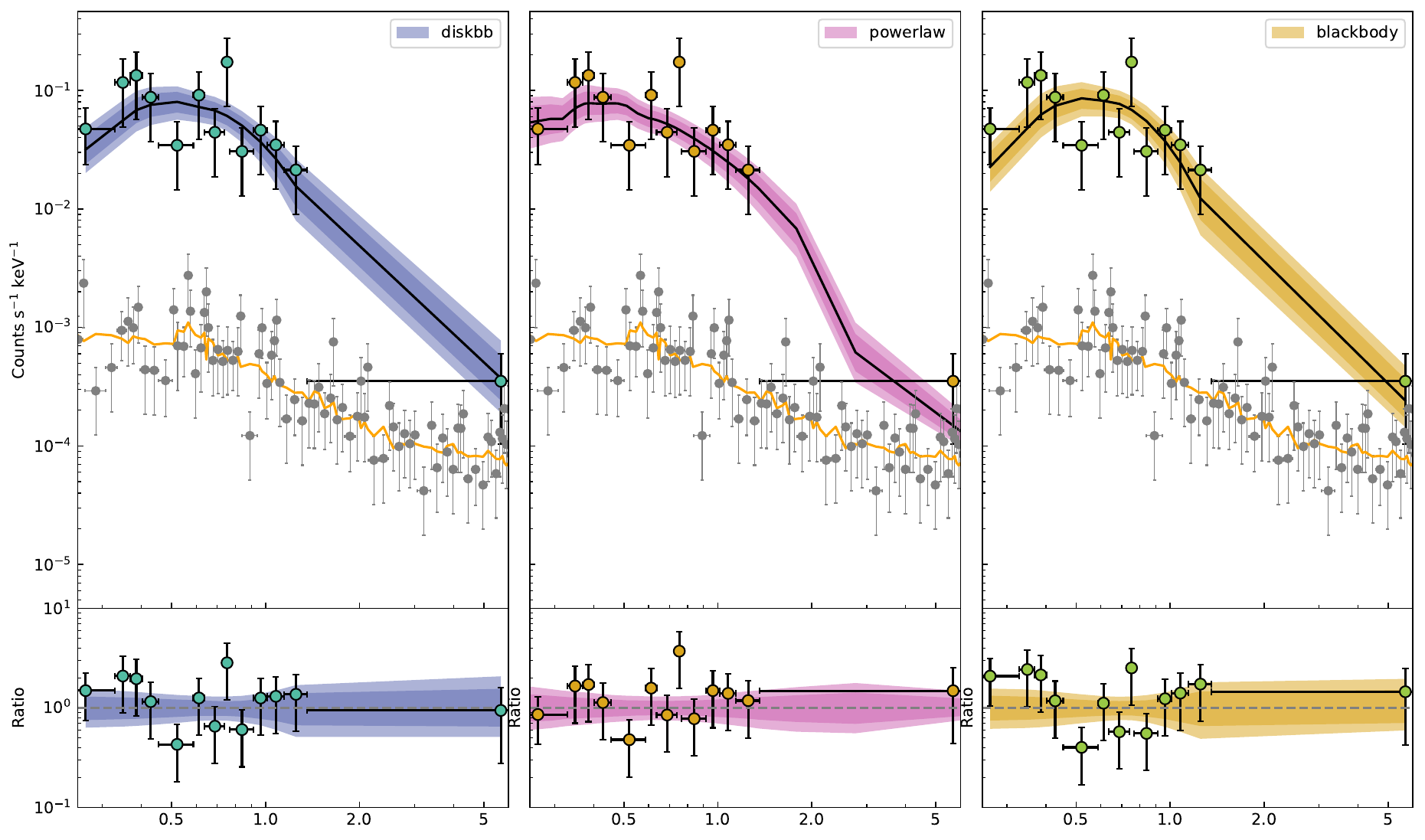}
\caption{eRASS:4 spectral fits and residuals. The grey dots and orange lines represent, respectively, the background rebinned data and model. \textit{Left:} diskbb model and uncertainties. \textit{center:} power-law model and uncertainties. \textit{center:} blackbody model and uncertainties.
}\label{eRASS_spec}
\end{figure}

\begin{figure}
    \centering
    \includegraphics[width=\textwidth]{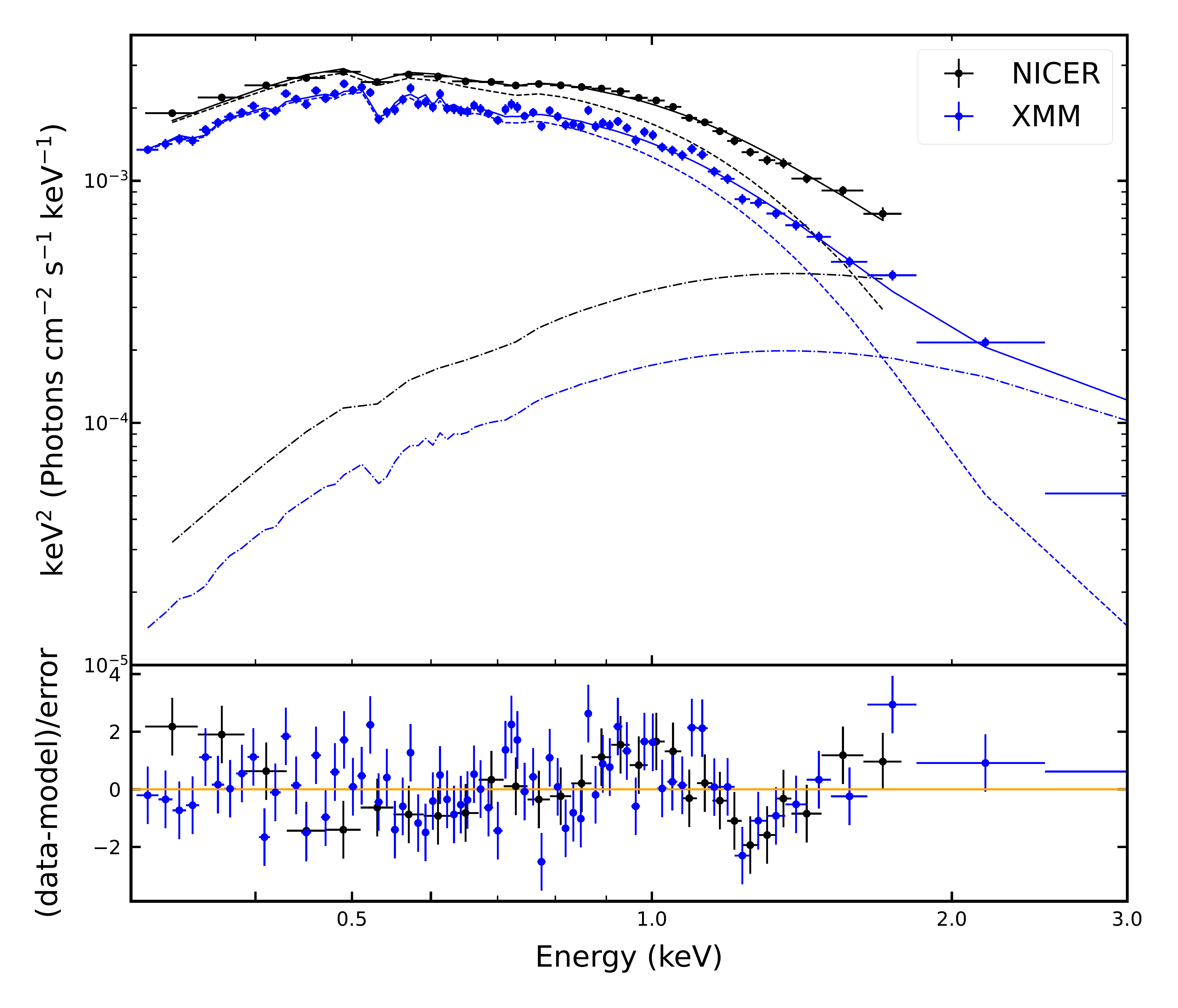}
    \caption{The {\xmm} pn spectrum (blue circles) and the NICER spectrum (black circles) fitted with {\tt tbabs*zxipcf*(diskbb+compTT).}}
\label{fig:xray_spec_xmm_nicer.png}
\end{figure}

\begin{figure}[h]%
\centering
\includegraphics[width=1\textwidth]{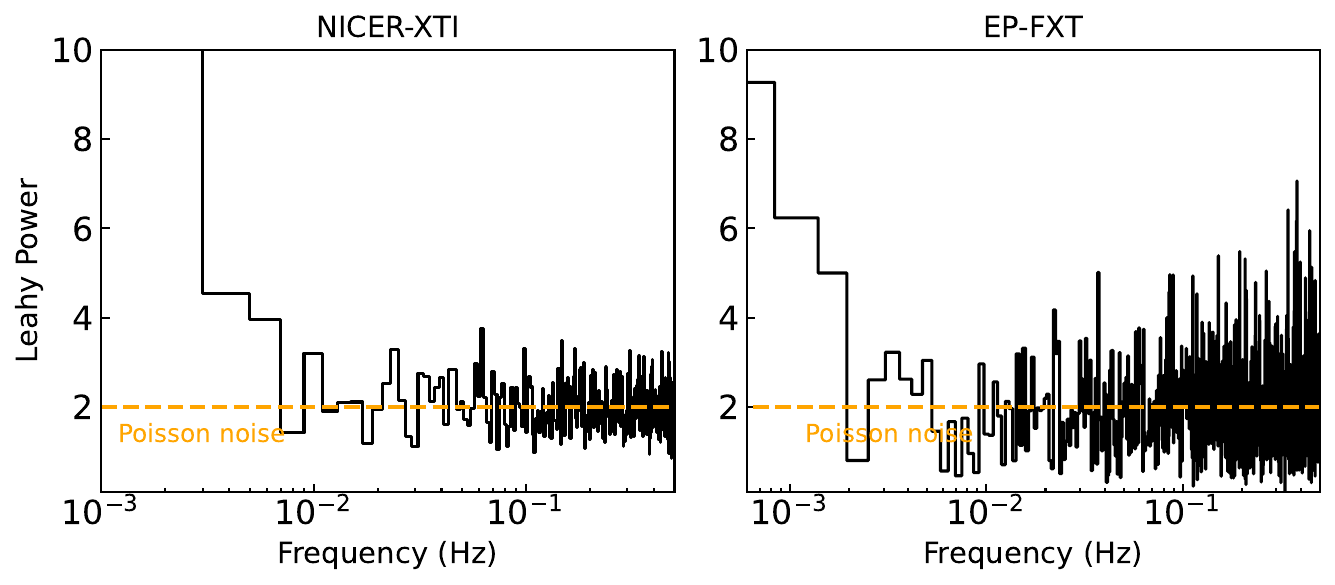}
\caption{EP240222a’s average X-ray PDS. {\bf a:} The average X-ray PDS from 7 continuous 500-s light curves taken with NICER-XTI. {\bf b:} The average X-ray PDS from 5 continuous 1800-s light curves taken with EP-FXT.}
\label{fig-pds}
\end{figure}

\begin{figure}[h]%
\centering
\includegraphics[width=0.9\textwidth]{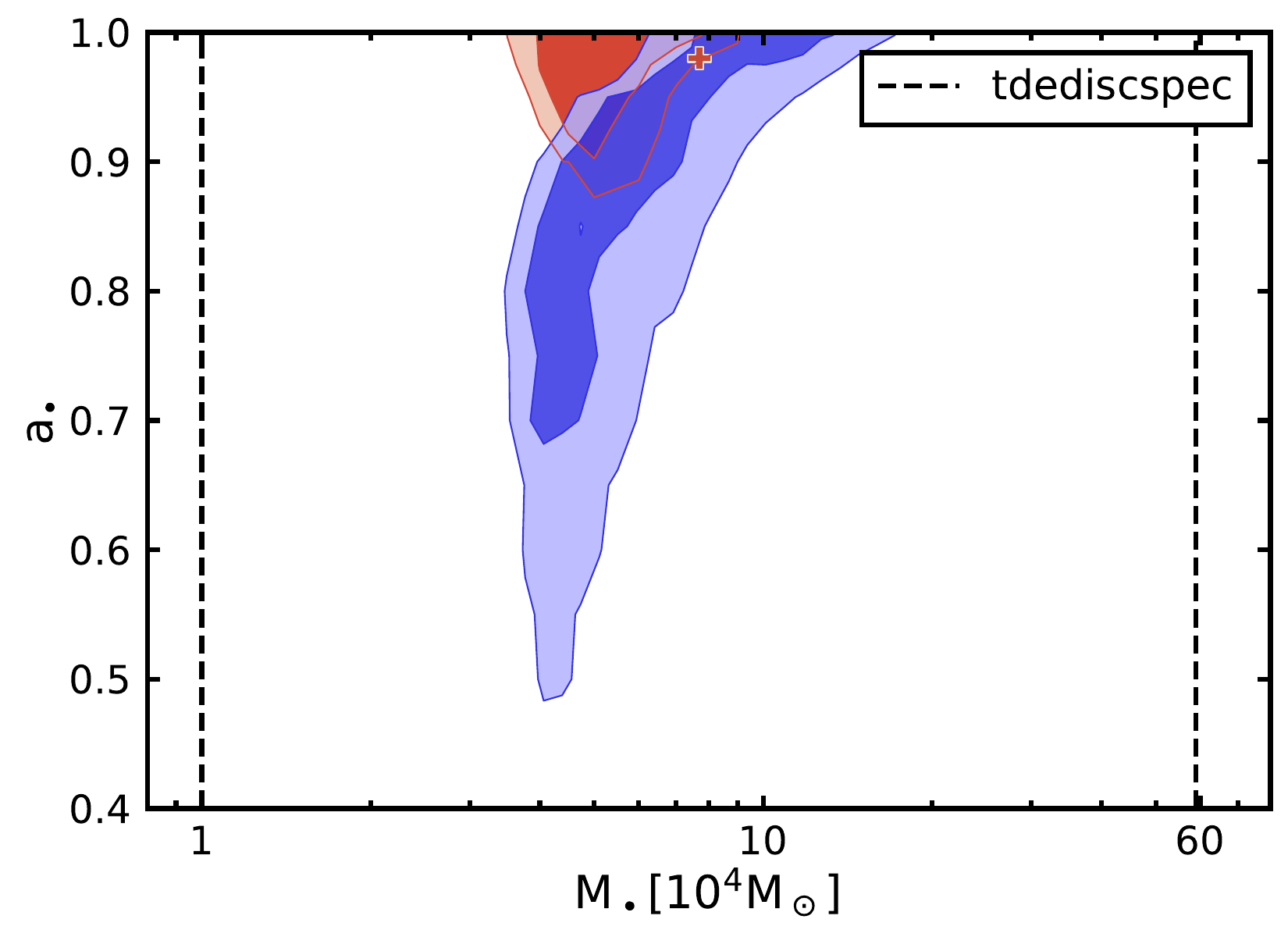}
\caption{Constraints on the black hole mass $M_\bullet$ and spin $a_\bullet$ from X-ray spectral fitting. We simultaneously fit the EP-FXT 2-4 epochs and the XMM epoch using the fit function {\tt TBabs$\times$THcomp$\times$slimd}. The red plus symbol indicates the best fit, while the blue and light blue regions represent the $1\sigma$ and $2\sigma$ confidence contours, respectively. The mass and spin were constrained to $M_\bullet = 7.7^{+4}_{-4} \times 10^4 M_\odot$ and $a_\bullet > 0.7$ at the $1\sigma$ confidence level, as determined by the slim disk modeling.}\label{ma}
\end{figure}

\begin{figure}[h]%
\centering
\includegraphics[width=\textwidth]{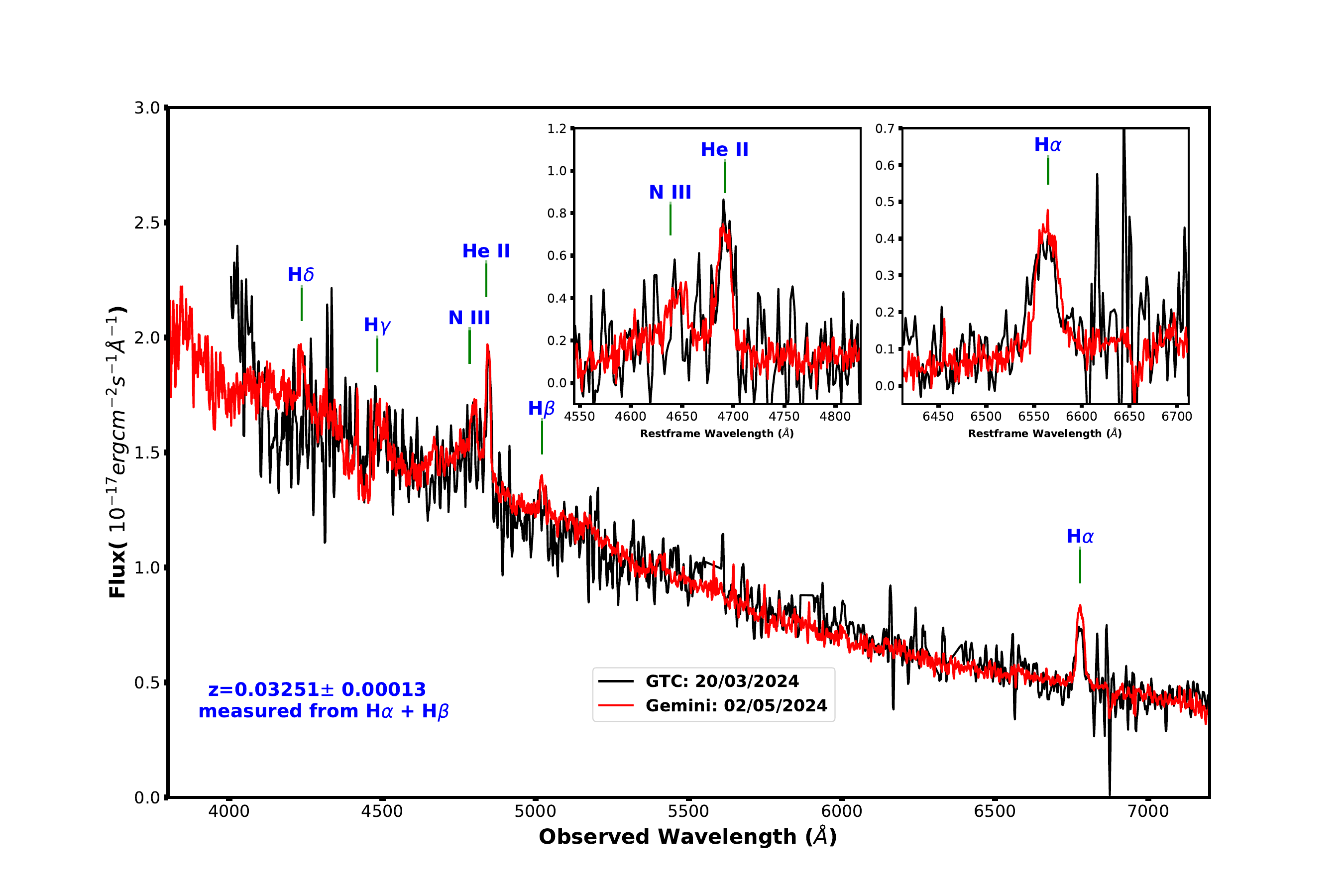} 
\caption{The GTC (black line) and Gemini (red line) spectrum of EP240222a. The spectrum exhibits He {\sc II} $\lambda$4686 and H$\alpha$ at a redshift of 0.032. The shape of the continuum is blue. Two insets in the figure display the features of He {\sc II} $\lambda$4686 and H$\alpha$. A single Gaussian model has been fitted to the lines, providing line width, with the luminosity calculated based on z=0.032.}
\label{fig:gtcspec}
\end{figure}

\begin{figure*}[!h]%
\centering
\includegraphics[width=0.9\textwidth]{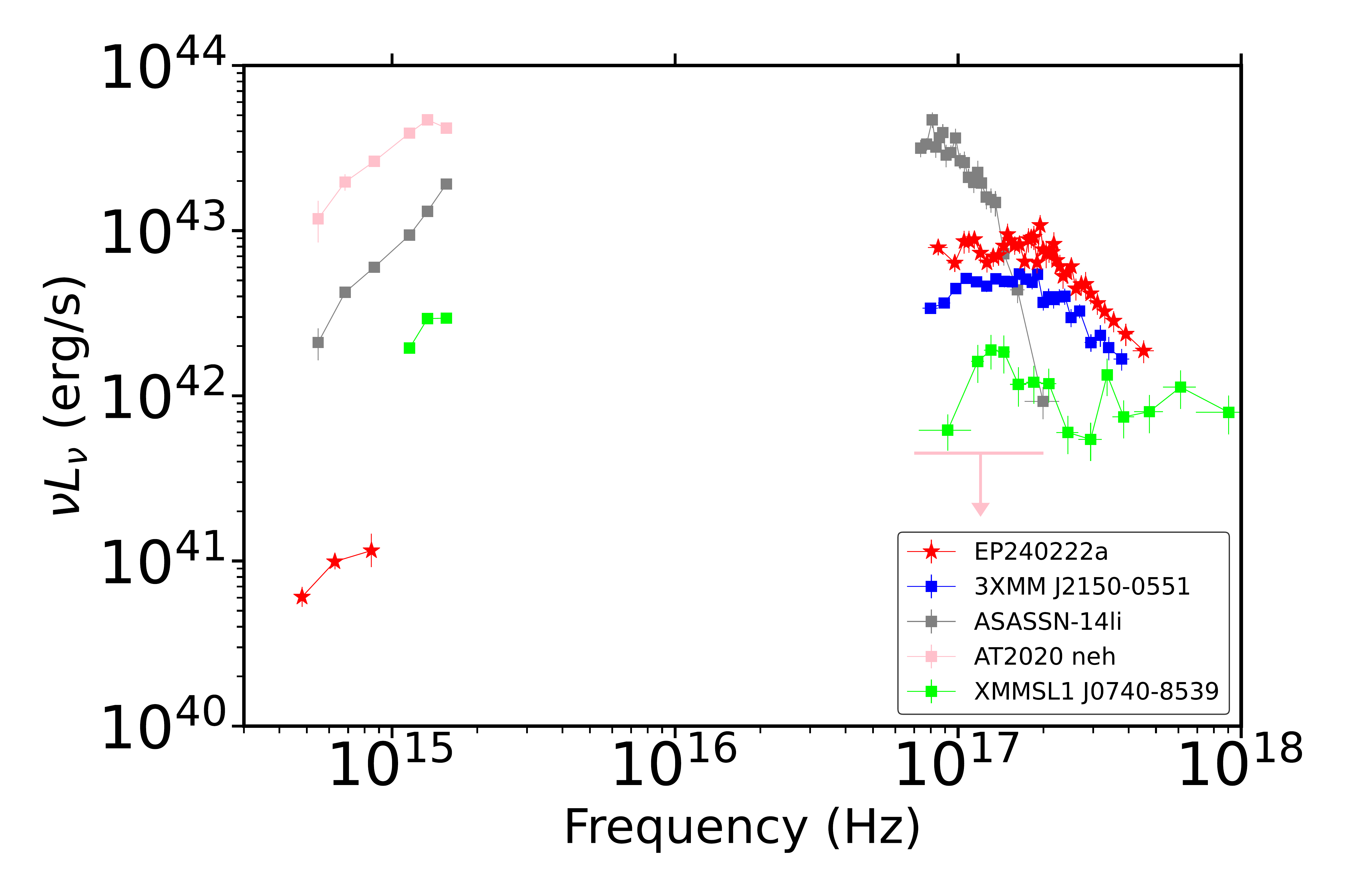}
\caption{Comparison between the broad band SED of EP240222a and some other TDEs. }
\label{fig:xray_opt_sed}
\end{figure*}

\begin{figure*}[!h]%
\centering
\includegraphics[width=0.9\textwidth]{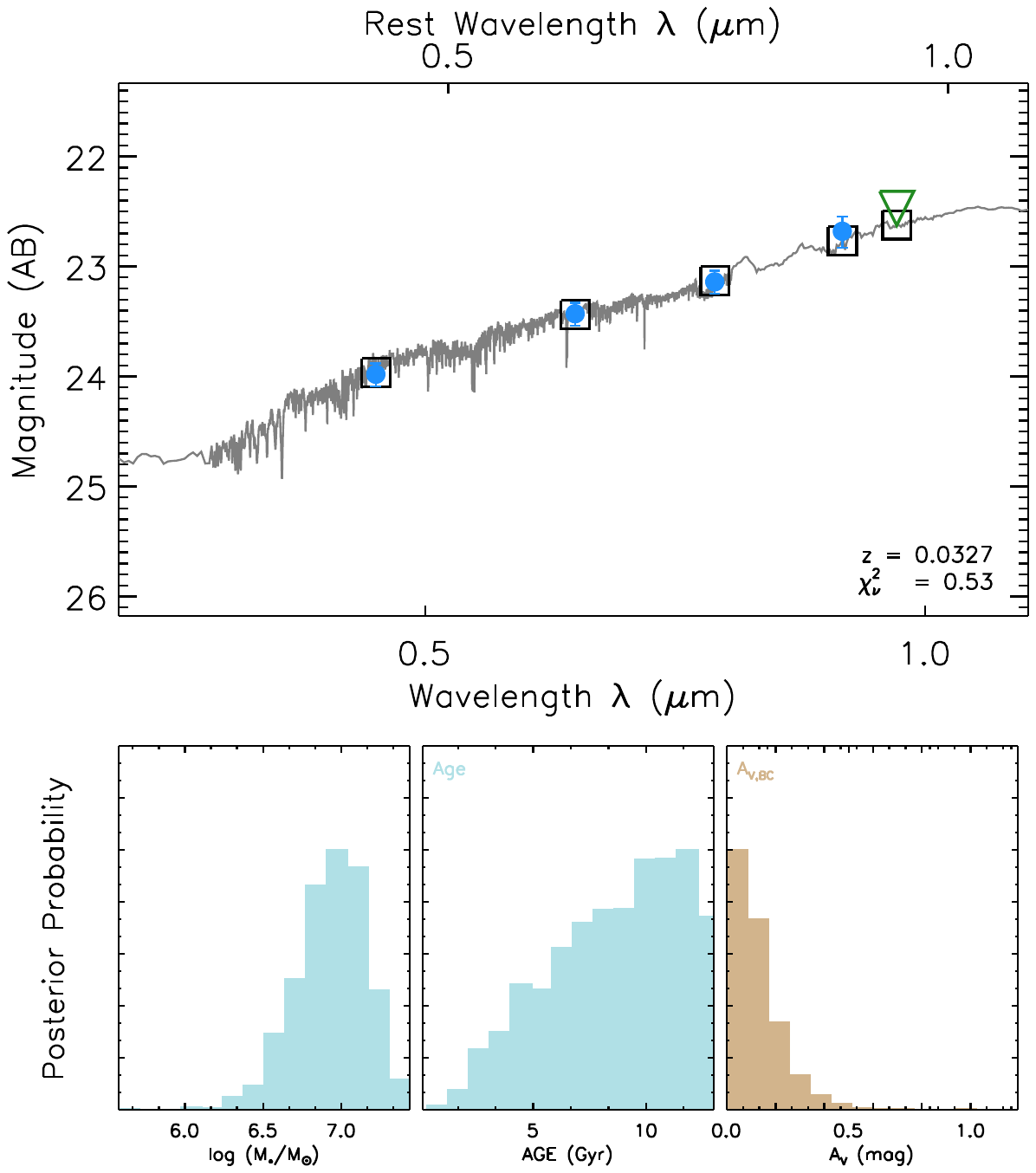}
\caption{Stellar population model fit to the optical photometric data  obtained before the outburst. The top panel shows the observed SED (blue circles), the best matched model (solid line) and the synthesis magnitudes (squares). Bottom panels present the posterior probabilities of stellar mass, age and extinction. }
\label{fig:hostfit}
\end{figure*}

\begin{figure}[h]%
\centering
\includegraphics[width=0.8\textwidth]{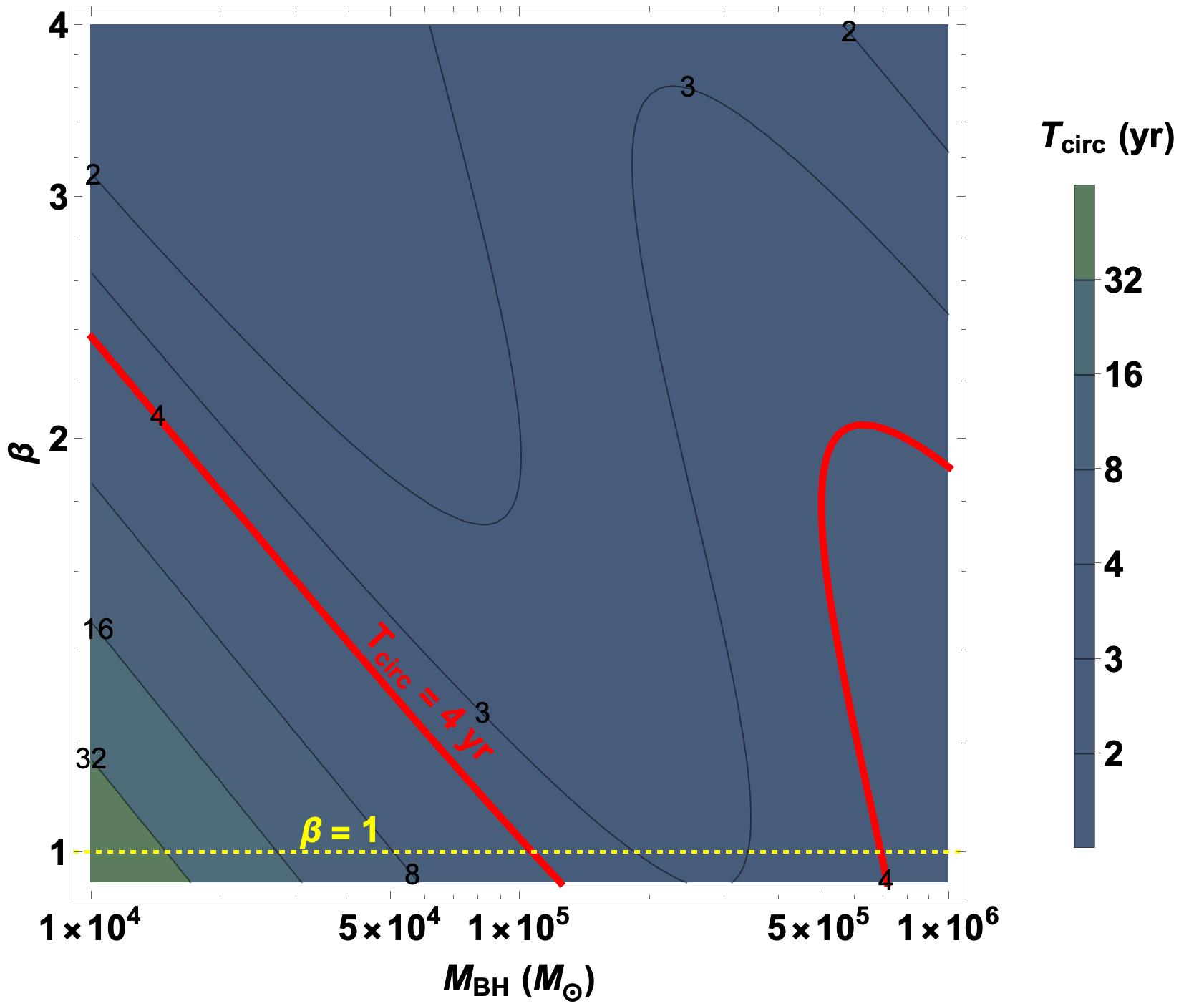}
\caption{The disk circularization timescale for a solar-type star: The x-axis is the black hole mass $M_{\rm BH}$. The y-axis is the stellar orbital penetration parameter $\beta$. The contours show the disk circularization timescale $T_{\rm circ}$ in units of years. The blue thick contour indicates the estimated disk formation time of EP240222a. The yellow dashed line marks $\beta=1$.}
\label{fig:Tcirc}
\end{figure}

\begin{figure*}[!h]%
\centering
\includegraphics[width=0.9\textwidth]{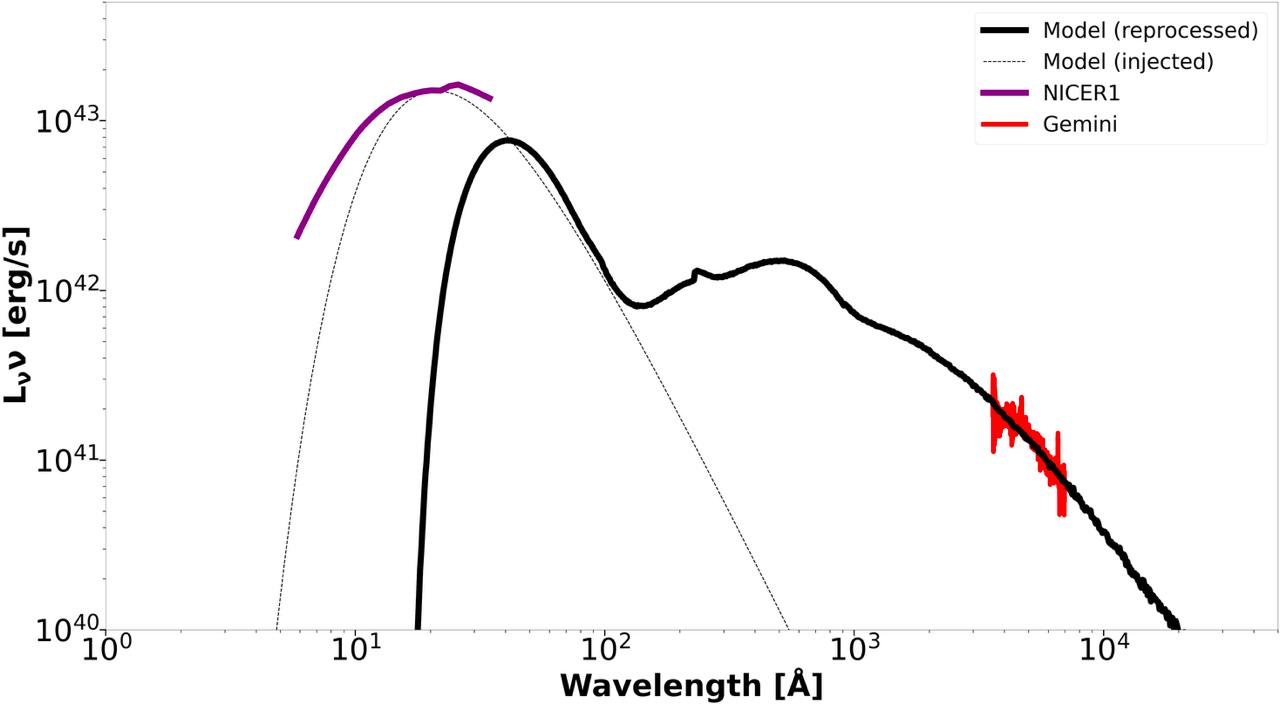}
\includegraphics[width=0.9\textwidth]{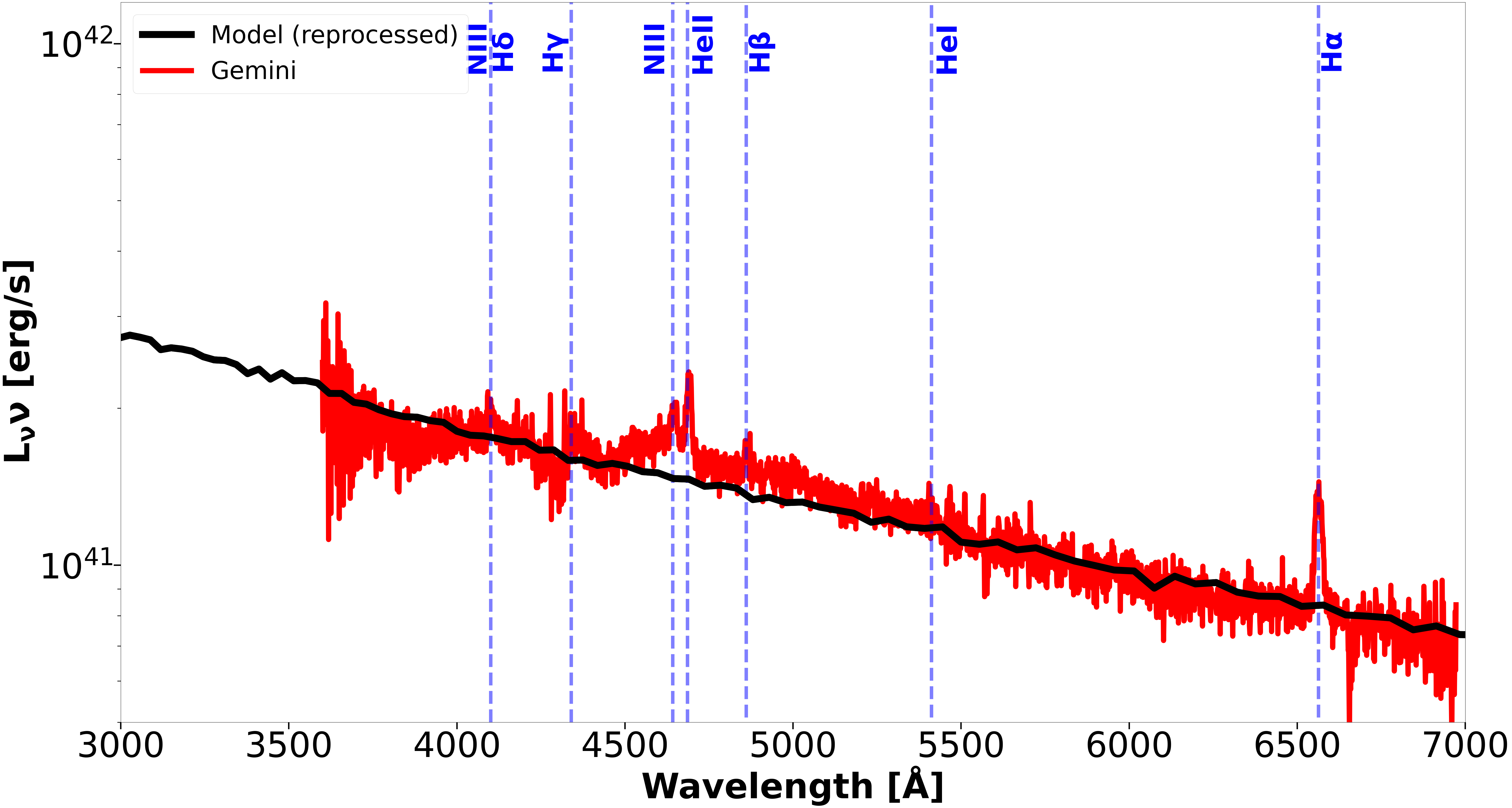}
\caption{Best-fit modelled spectrum in comparison to various observations. The model parameters are $M_{\rm BH}=10^5 M_\odot$, $M_{\rm env} = 0.002 M_\odot$, $q = 0.5$,  $L_X =1.5 \times 10^{43} \rm erg \ s^{-1}$ and $v_0=0.01c$. The top panel displays the complete simulated and observed SED. The spectrum escaping the envelope in the simulation is represented by a black solid curve and the injected spectrum is shown as a black dotted curve. We compare the model spectra to the real observations: the X-ray spectrum observed by NICER on 14 March 2024, and the optical spectrum observed by Gemini on 2 May 2024. For a closer look at the optical bands, a zoomed-in comparison is shown in the lower panel. 
}
\label{fig:Best_fit_SED}
\end{figure*}

\clearpage


\begin{longtable}{cccc}
\caption{X-ray observational log of EP240222a.} \label{tab:phot} \\
\hline \multicolumn{1}{c}{Observational time} & \multicolumn{1}{c}{obsid} & \multicolumn{1}{c}{exposure(s)} & 
\multicolumn{1}{c}{${F_{\rm unabs, 0.5-4 keV}}$} 
\endfirsthead
\multicolumn{4}{c}%
{{\bfseries \tablename\ \thetable{} -- continued from previous page}} \\
\hline \multicolumn{1}{c}{Observational time} & \multicolumn{1}{c}{obsid} & \multicolumn{1}{c}{exposure(s)} & 
\multicolumn{1}{c}{${F_{\rm unabs, 0.5-4 keV}}$} \\ \hline 
\endhead
\hline \multicolumn{4}{r}{{Continued on next page}} \\
\endfoot
\hline
\multicolumn{4}{l}{}
\endlastfoot    
\hline
& SRG/eROSITA & \\
2020-05-23T02:14:27 & eRASS1 & 166 & $< -12.65$  \\ 
2020-11-22T03:44:45 & eRASS2 & 190 & $-12.90\pm0.19$ \\
2021-05-23T07:14:26 & eRASS3 & 165 & $-13.02\pm0.22$ \\ 
2021-11-24T08:44:42 & eRASS4 & 160 & $-13.09\pm0.20$ \\
& eRASS:4 & 680 & $-13.06\pm0.12$  \\ 
\hline
& LEIA & \\
2024-01-31T19:23:59 & 06800008449 & 949 & - \\
2024-01-31T20:57:46 & 06800008450 & 941 & -\\
2024-01-31T22:31:32 & 06800008451 & 943 & -\\
2024-02-01T00:05:18 & 06800008452 & 742 & -\\
stacked LEIA        &      -      &3575 & $-11.1_{-0.2}^{+0.2}$ \\
\hline
& EP-WXT & \\
2024-02-22T07:00:46 & 08500000008 & 21912  & $-11.1_{-0.1}^{+0.1}$\\ 
2024-03-11T08:26:47 & 13600005107 & 16845  & $-10.9_{-0.2}^{+0.2}$\\
2024-03-12T00:29:53 & 13600005115 & 12967  & $-11.4_{-0.2}^{+0.2}$\\
2024-03-25T11:33:22 & 08503014656 & 38432  & $-11.0_{-0.2}^{+0.2}$\\
2024-03-28T19:28:50 & 11911403264 & 65133  & $-11.3_{-0.2}^{+0.2}$\\
2024-03-31T13:19:11 & 11912597761 & 38767  & $-10.6_{-0.2}^{+0.2}$\\
2024-04-04T20:00:55 & 11904194446 & 9261   & $-11.2_{-0.2}^{+0.2}$\\
\hline      
& EP-FXT & \\
2024-03-13T00:33:29 & 08500000022 & 1960  & $-11.10_{-0.04}^{+0.2}$\\ 
2024-04-17T11:31:30 & 08500000070 & 11881 & $-11.48_{-0.01}^{+0.01}$\\  
2024-05-23T16:02:50 & 08500000100 & 35919 & $-11.586_{-0.006}^{+0.006}$\\
\hline
& Chandra/HRC & \\
2024-04-01T14:01:12 & 29356 & 1991 & -10.99$_{-0.02}^{+0.02}$\\ 
\hline
& {\xmm} & \\
2024-05-23T19:28:52 & 0923470101 & 54575 & $-11.503_{-0.005}^{+0.007}$\\ 
\hline
& Nicer/XTI & \\
2024-03-14T00:04:32 & 7204250101 & 2729 & $-11.19_{-0.03}^{+0.02}$ \\
2024-03-15T02:24:27 & 7204250102 & 725  & $-11.19_{-0.08}^{+0.07}$ \\
2024-03-16T04:46:31 & 7204250103 & 2029 & $-11.18_{-0.03}^{+0.02}$ \\
2024-03-17T01:24:40 & 7204250104 & 1387 & $-11.23_{-0.02}^{+0.03}$ \\
2024-03-18T04:48:09 & 7204250105 & 512  & $-11.18_{-0.05}^{+0.04}$ \\
2024-03-21T04:04:10 & 7204250107 & 1832 & $-11.21_{-0.04}^{+0.03}$ \\
2024-03-22T06:24:04 & 7204250108 & 1959 & $-11.24_{-0.04}^{+0.01}$ \\
2024-03-23T07:12:04 & 7204250109 & 1256 & $-11.29_{-0.02}^{+0.03}$  \\
2024-03-24T06:26:04 & 7204250110 & 979  & $-11.19_{-0.04}^{+0.04}$ \\
2024-03-26T06:34:20 & 7204250111 & 276  & $-11.24_{-0.04}^{+0.13}$ \\
2024-03-27T02:37:15 & 7204250112 & 1887 & $-11.22_{-0.03}^{+0.02}$ \\ 
2024-03-28T04:53:21 & 7204250113 & 2012 & $-11.26_{-0.02}^{+0.02}$ \\
2024-04-05T00:17:00 & 7204250116 & 186  & $-11.32_{-0.03}^{+0.13}$ \\
2024-04-06T01:04:28 & 7204250117 & 144  & $-11.31_{-0.05}^{+0.05}$ \\
2024-04-07T00:18:06 & 7204250118 & 80   & $-11.10_{-0.20}^{+0.24}$  \\
2024-04-09T00:32:49 & 7204250120 & 1920 & $-11.27_{-0.02}^{+0.03}$ \\
2024-04-19T20:57:00 & 7204250121 & 154  & $-11.58_{-0.05}^{+0.07}$ \\
2024-04-19T23:56:40 & 7204250122 & 2036 & $-11.34_{-0.15}^{+0.08}$ \\
2024-04-21T08:28:40 & 7204250123 & 490  & $-11.32_{-0.05}^{+0.05}$ \\
2024-04-21T23:49:00 & 7204250124 & 804  & $-11.36_{-0.02}^{+0.11}$ \\
2024-04-23T20:41:00 & 7204250125 & 346  & $-11.30_{-0.11}^{+0.16}$ \\
2024-04-26T15:41:20 & 7204250126 & 1025 & $-11.39_{-0.03}^{+0.05}$ \\
2024-04-28T15:38:20 & 7204250127 & 486  & $-11.34_{-0.05}^{+0.07}$ \\
2024-04-29T16:07:00 & 7204250128 & 2921 & $-11.42_{-0.03}^{+0.03}$ \\
2024-05-02T06:22:11 & 7204250129 & 1488 & $-11.48_{-0.06}^{+0.06}$ \\
2024-05-04T12:15:42 & 7204250130 & 1627 &
$-11.55_{-0.05}^{+0.04}$\\
2024-05-05T00:38:43 & 7204250131 & 3396 & $-11.39_{-0.03}^{+0.04}$ \\
2024-05-06T01:17:41 & 7204250132 & 6001 & $-11.50_{-0.02}^{+0.07}$ \\
2024-05-07T00:42:06 & 7204250133 & 3735 & $-11.51_{-0.07}^{+0.05}$ \\
2024-05-08T01:30:54 & 7204250134 & 625  & $-11.38_{-0.02}^{+0.08}$ \\
2024-05-23T09:49:06 & 7204250136 & 2051 & $-11.58_{-0.03}^{+0.01}$ \\
2024-05-29T09:44:11 & 7204250137 & 2661 & $-11.49_{-0.03}^{+0.06}$
\label{tab:xray_label}
\end{longtable}

\clearpage

\begin{table}[]
    \centering
    \caption{Best-fit values and the 1$\sigma$ errors on the spectral fitting to the {\xmm} and averaged NICER spectrum with {\tt zxipcf*(diskbb+compTT)}. Both fits include absorption by Galactic column (model {\tt tbabs}, $n_{\rm H} = 1.8 \times 10 ^{20}$ atoms cm$^{-2}$, abundances from \cite{Wilms2000}). The redshift of {\tt zxipcf} component was fixed to the value of 0.032, and the approx of {\tt compTT} was fixed to 1 during the fitting.}
    \begin{tabular}{ccccccc}
    \hline
    spectrum  &  $N_{\rm H}$ & log$\xi$ & Tin &  kT & $\tau$ & $C (d.o.f)$\\
              & (10$^{20}$\,cm$^{-2}$) & (erg\,s$^{-1}$\,cm) & (eV) & (eV) &\\  
    \hline
    {\xmm}       & 6.8$^{+5.1}_{-0.7}$ & -0.7$^{+0.2}_{-0.6}$ & 181$^{+5}_{-15}$ & 595$^{+268}_{-98}$& 9.1$^{+1.2}_{-1.2}$ & 310 (287)\\
    \hline
    NICER1    & 14.2$^{+13.6}_{-10.7}$ & -1.4$^{+2.0}_{-1.6}$ & 216$^{+22}_{-47}$ & 656$^{+*}_{-331}$ & 14.2$^{+*}_{-*}$ & 23.39(34) \\
    \hline
    \end{tabular} 
\newline
\footnotesize
{\bf Notes:} The errors of these parameters can not be well constrained during the spectral fitting.
    \label{tab:xmm_fitting}
\end{table}

\begin{table}
    \centering
    \caption{Best-fit values and $1\sigma$ errors for the fit parameters are presented. EP1 was fitted separately, while the spectra from the other epochs were fitted simultaneously using the models TBabs$\times$THcomp$\times$slimdisk and TBabs$\times$THcomp$\times$tdediscspec. $^a$: The accretion rate $\dot{m}=\dot{M}/\dot{M}_{\rm Edd}$ was estimated by fixing $M_\bullet$ and $a_\bullet$ at their best-fit values. The Eddington mass accretion rate is defined as $\dot M_{\rm Edd} = 1.37 \times 10^{19} M_4~{\rm kg~s}^{-1}$. $^b$: The black hole mass $M_\bullet$ was estimated from $R_p$.}
    \begin{tabular}{cc|c|cccc}
    \hline
    Model &Parameter & EP1  & EP2  & EP3& EP4 &XMM \\
     \hline
    TBabs&$N_{\rm H}$ ($10^{20} {\rm cm^{-2}}$)  &$3.3^{+4.4}_{-2.3}$ & $4.5\pm0.7$ &=EP2&=EP2& $3.5^{+0.6}_{-0.7}$ \\
    THcomp& $\tau$&-&$0.5\pm0.3$&=EP2&=EP2&$0.5^{+0.8}_{-0.3}$\\
    & $kT_e$ (keV)&-&$30^{+30}_{-12}$&=EP2&=EP2&=EP2\\
    slim disk& $M_\bullet$ ($10^4M_\odot$) &$5^{+3}_{-1}$&$7.7^{+4.0}_{-4.0}$&=EP2&=EP2&=EP2 \\
    &$a_\bullet$   &$0.998_{-0.1}$&$0.98^{+0.02}_{-0.3}$&=EP2&=EP2&=EP2 \\
    
    &$\dot m$ (Edd)$^a$ &$10^{+90}_{-3}$&$11^{+12}_{-7}$&$5.5^{+2.4}_{-2.7}$&$3.5^{+1.1}_{-1.3}$&=EP3 \\
    &$ \theta$ (deg)  &$10^{+23}_{-8}$&$60^{+6}_{-58}$&=EP2&=EP2&=EP2 \\
    &$ f_c$  &$2.88^{+0.05}_{-0.09}$&-&-&-&- \\
    & Cstat/$\nu$  &$355.9/393$&$297.9/259$&$281.0/306$&$147.1/164$&$681.5/613$\\
    \hline
    TBabs&$N_{\rm H}$ ($10^{20} {\rm cm^{-2}}$)  &$3.4\pm0.4$&$4.5\pm0.1$&=EP2&=EP2& $3.7\pm0.1$\\
    THcomp& $\tau$&-&$0.25\pm0.04$&=EP2&=EP2&$0.27\pm0.01$\\
    & $kT_e$ (keV)&-&$62\pm1$&=EP2&=EP2&=EP2\\
    tdediscspec 
      & $R_p$ ($10^{10}$cm)  
       &$2.6\pm0.4$&$6.3\pm0.6$&=EP2&=EP2&=EP2\\
    & $M_\bullet$ ($10^4M_\odot$) $^b$ &$13\pm12$&$31\pm28$&=EP2&=EP2&=EP2 \\
       & $T_{\rm p}$ ($10^6$K)  &$2.89\pm0.01$&$1.73\pm0.01$&$1.68\pm0.01$&$1.64\pm0.01$& =EP3 \\
       & $\gamma$  &$1.5_{-0.02}$&$1.27\pm0.01$&=EP2&=EP2&=EP2\\
      & Cstat/$\nu$  &$355.3/393$&$303.0/259$&$282.1/306$&$145.3/164$&$682.5/613$\\
        \hline
    \end{tabular}
    \label{tab:Xrayspfit}
\end{table}

\clearpage


\begin{longtable}{ccccc}
\caption{Optical observational log of EP240222a. $^a$ The source was not detected during these observations, and the limit magnitudes are listed here.$^b$ The detection signal to noise ratio is less than 3.} 
\label{tab:phot}\\
\hline 
\multicolumn{1}{c}{Instrument} & \multicolumn{1}{c}{Date (UT)} & \multicolumn{1}{c}{Filter} & 
\multicolumn{1}{c}{Exposure time (s)} & \multicolumn{1}{c}{Magnitude} 
\endfirsthead
\multicolumn{5}{c}%
{{\textbf{\tablename\ \thetable{} -- continued from previous page}}} \\
\hline 
\multicolumn{1}{c}{Instrument} & \multicolumn{1}{c}{Date (UT)} & \multicolumn{1}{c}{Filter} & 
\multicolumn{1}{c}{Exposure time (s)} & \multicolumn{1}{c}{Magnitude}\\
\hline
\endhead
\hline 
\multicolumn{5}{r}{{Continued on next page}} \\
\endfoot
\multicolumn{5}{l}{}
\endlastfoot 
\hline
    Xinglong-Schmidt &    2024-03-12  & $r$ & 2400 & 21.2 $\pm$ 0.2 \\ 
	Xinglong-2.16	&    2024-03-13   & clear & 2760 &21.7 $\pm$ 0.5 \\ 
	Xinglong-2.16	&    2024-03-14   & clear & 900  & 22.0$^a$        \\ 
	Xinglong-2.16	&    2024-03-19   & clear & 1200 & 21.5$^a$ \\
	WFST  &  2024-03-14  & $u$  &  480  &  21.31$^b$ $\pm$ 0.32  \\
    WFST  &  2024-03-14  & $g$  &  360  &  21.23 $\pm$ 0.20  \\
    WFST  &  2024-03-17  & $g$  &  360  &  21.39$^b$ $\pm$ 0.36  \\
    WFST  &  2024-03-17  & $r$  &  360  &  21.65 $\pm$ 0.21  \\
    WFST  &  2024-03-19  & $u$  &  360  &  21.33 $\pm$ 0.25  \\
    WFST  &  2024-03-19  & $g$  &  360  &  21.28 $\pm$ 0.08  \\
    WFST  &  2024-03-19  & $r$  &  360  &  21.55 $\pm$ 0.17  \\
    WFST  &  2024-03-28  & $g$  &  360  &  21.30 $\pm$ 0.08  \\
    WFST  &  2024-03-28  & $r$  &  360  &  21.55 $\pm$ 0.17  \\
    WFST  &  2024-03-29  & $r$  &  360  &  21.57 $\pm$ 0.11  \\
    WFST  &  2024-03-30  & $g$  &  360  &  21.37 $\pm$ 0.06  \\
    WFST  &  2024-03-31  & $r$  &  360  &  21.58 $\pm$ 0.14  \\
    WFST  &  2024-04-01  & $g$  &  360  &  21.34 $\pm$ 0.08  \\
    WFST  &  2024-04-01  & $r$  &  360  &  21.67 $\pm$ 0.24  \\
    WFST  &  2024-04-02  & $g$  &  360  &  21.30 $\pm$ 0.06  \\
    WFST  &  2024-04-02  & $r$  &  360  &  21.45 $\pm$ 0.14  \\
    WFST  &  2024-04-03  & $r$  &  360  &  21.60 $\pm$ 0.08  \\
    WFST  &  2024-04-05  & $g$  &  360  &  21.42 $\pm$ 0.06  \\
    WFST  &  2024-04-05  & $r$  &  360  &  21.56 $\pm$ 0.07  \\
    WFST  &  2024-04-06  & $r$  &  360  &  21.55 $\pm$ 0.05  \\
    WFST  &  2024-04-07  & $g$  &  300  &  21.31 $\pm$ 0.05  \\
    WFST  &  2024-04-07  & $r$  &  300  &  21.70 $\pm$ 0.11  \\
    WFST  &  2024-04-09  & $g$  &  300  &  21.35 $\pm$ 0.05  \\
    WFST  &  2024-04-09  & $r$  &  300  &  21.62 $\pm$ 0.06  \\
    WFST  &  2024-04-10  & $g$  &  300  &  21.37 $\pm$ 0.05  \\
    WFST  &  2024-04-10  & $r$  &  300  &  21.51 $\pm$ 0.08  \\
    WFST  &  2024-04-11  & $r$  &  300  &  21.55 $\pm$ 0.09  \\
    WFST  &  2024-04-13  & $r$  &  300  &  21.78 $\pm$ 0.13  \\
    WFST  &  2024-04-15  & $g$  &  300  &  21.34 $\pm$ 0.07  \\
    WFST  &  2024-04-15  & $r$  &  300  &  21.74 $\pm$ 0.14  \\
    WFST  &  2024-05-03  & $r$  &  120  &  21.62 $\pm$ 0.21  \\
    WFST  &  2024-05-04  & $g$  &  120  &  21.71 $\pm$ 0.22  \\
    WFST  &  2024-05-06  & $g$  &  120  &  21.54 $\pm$ 0.20  \\
    WFST  &  2024-05-07  & $r$  &  120  &  22.10 $\pm$ 0.29  \\
    WFST  &  2024-05-27  & $g$  &  240  &  21.65$^b$ $\pm$ 0.30  \\
    WFST  &  2024-06-13  & $g$  &  180  &  22.11$^b$ $\pm$ 0.58  \\
    WFST  &  2024-06-21  & $g$  &  120  &  21.57$^b$ $\pm$ 0.31  \\
    WFST  &  2024-06-22  & $r$  &  150  &  21.75$^b$ $\pm$ 0.52  \\
    WFST  &  2024-06-28  & $r$  &  180  &  22.00$^b$ $\pm$ 0.33  \\
    WFST  &  2024-06-30  & $g$  &  180  &  22.45$^b$ $\pm$ 0.42  \\
    WFST  &  2024-07-04  & $g$  &  180  &  21.40 $\pm$ 0.27  \\
    WFST  &  2024-07-07  & $g$  &  180  &  21.71 $\pm$ 0.13  \\
  \hline
	\label{table:catalog_opt}
\footnotesize
\newline
\end{longtable}

\clearpage


\clearpage 




\end{document}